\documentclass[reqno,10pt,a4paper]{amsart}

\numberwithin{equation}{section}
\allowdisplaybreaks

\usepackage{amssymb,eucal,mathrsfs,setspace,cite}
\usepackage[inner=20mm, outer=20mm, top=24mm, bottom=24mm, head=10mm, foot=10mm]{geometry} 
\usepackage[dvipsnames]{xcolor}

\usepackage{enumitem}
\setitemize{leftmargin=*}                
\setenumerate{leftmargin=*,              
              label=\textup{(\arabic*)}} 

\usepackage{array}
\newcolumntype{C}{>{$}c<{$}} 

\usepackage{mathptmx}                            

\usepackage{mathtools} 
\usepackage{stmaryrd} 

\usepackage[capitalise,noabbrev]{cleveref} 

\usepackage{tikz} 
\usetikzlibrary{shapes}

\newcommand{\fld}[1]{\mathbb{#1}}    
\newcommand{\alg}[1]{\mathfrak{#1}}  
\newcommand{\assalg}[1]{\mathrm{#1}} 
\newcommand{\grp}[1]{\mathsf{#1}}    
\newcommand{\Mod}[1]{\mathcal{#1}}   
\newcommand{\VOA}[1]{\mathsf{#1}}    
\newcommand{\poly}[1]{\mathsf{#1}}   
\renewcommand{\vec}[1]{\mathbf{#1}}    

\newcommand{\dd}{\mathrm{d}}   
\newcommand{\ii}{\mathfrak{i}} 
\newcommand{\ee}{\mathsf{e}}   

\newcommand{\wun}{\mathsf{1}}

\newcommand{\Walg}[1]{W$_{\! #1}$} 

\renewcommand{\ge}{\geq} 
\renewcommand{\le}{\leq} 

\DeclarePairedDelimiter{\brac}{\lparen}{\rparen}   
\DeclarePairedDelimiter{\sqbrac}{\lbrack}{\rbrack} 
\DeclarePairedDelimiter{\set}{\lbrace}{\rbrace}
\newcommand{\st}{\mspace{5mu} : \mspace{5mu}}      
\DeclarePairedDelimiter{\ang}{\langle}{\rangle}    

\DeclarePairedDelimiter{\normord}{{:}}{{:}}        

\DeclarePairedDelimiterX{\comm}[2]{\lbrack}{\rbrack}{#1 , #2}                            
\DeclarePairedDelimiterX{\acomm}[2]{\lbrace}{\rbrace}{#1 , #2}                           
\DeclarePairedDelimiterX{\super}[2]{\lparen}{\rparen}{#1 \delimsize\vert \mathopen{} #2} 

\newcommand{\ZZ}{\fld{Z}}
\newcommand{\QQ}{\fld{Q}}

\newcommand{\CC}{\fld{C}}

\newcommand{\symgp}[1]{\grp{S}_{#1}}   

\newcommand{\affine}[1]{\widehat{#1}}
\newcommand{\SLA}[2]{\alg{#1}\brac*{#2}}                     
\newcommand{\AKMA}[2]{\affine{\alg{#1}}\brac*{#2}}           
\newcommand{\UEA}[1]{\assalg{U} (#1)}                   

\newcommand{\hvs}{\alg{h}}                              
\newcommand{\dual}{\ast}                                
\newcommand{\dvs}{\hvs^\dual}                           
\newcommand{\kil}[2]{\brac*{#1,#2}}                     
\newcommand{\heis}{\widehat{\hvs}}                      
\newcommand{\gae}[1]{\ee^{#1}}                          
\newcommand{\scrt}[1]{a^{#1}}                           
\newcommand{\dscrt}[1]{a^{\dual #1}}                    
\newcommand{\fwt}[1]{\omega_{#1}}                       
\newcommand{\srt}[1]{\alpha^{#1}}                       

\newcommand{\weyl}{\varrho}                             

\newcommand{\hvoa}[1]{\VOA{H}_{#1}}                     
\newcommand{\Wvoa}[1]{\VOA{W}_{#1}}                     
\newcommand{\wvoa}{\Wvoa{3}}                            
\newcommand{\wminmod}[1]{\wvoa(#1)}                     

\newcommand{\Fock}[1]{\Mod{F}_{#1}}                     
\newcommand{\svwts}[2]{\zeta_{#1;#2}}                   
\newcommand{\Verma}[1]{\Mod{V}_{#1}}                    

\newcommand{\lra}{\longrightarrow}
\newcommand{\ira}{\hookrightarrow}                                        

\newcommand{\blank}{{-}} 

\DeclarePairedDelimiter{\ket}{\lvert}{\rangle}
\DeclarePairedDelimiterX{\braket}[2]{\langle}{\rangle}{#1 \delimsize\vert \mathopen{} #2}
\DeclarePairedDelimiterX{\bracket}[3]{\langle}{\rangle}{#1 \delimsize\vert \mathopen{} #2 \delimsize\vert \mathopen{} #3}

\newcommand{\partn}[1]{\sqbrac*{#1}}                 
\newcommand{\len}[1]{\ell\brac*{#1}}                 

\newcommand{\sym}{\Lambda}                                       
\newcommand{\fsym}[1]{\sym_{#1}}                                 

\newcommand{\monsym}[1]{\poly{m}_{#1}}                           
\newcommand{\fmonsym}[2]{\poly{m}_{#1} \brac[\big]{#2}}          

\newcommand{\powsum}[1]{\poly{p}_{#1}}                           
\newcommand{\fpowsum}[2]{\poly{p}_{#1} \brac[\big]{#2}}
\newcommand{\jack}[2]{\poly{P}_{#1}^{#2}}                        
\newcommand{\fjack}[3]{\poly{P}_{#1}^{#2} \brac[\big]{#3}}
\newcommand{\djack}[2]{\poly{Q}_{#1}^{#2}}                       
\newcommand{\fdjack}[3]{\poly{Q}_{#1}^{#2} \brac[\big]{#3}}

\newcommand{\nc}[2]{b_{#1}^{#2}}                     
\newcommand{\fintker}[3]{G_{#1}^{#2}(#3)}
\newcommand{\jprod}[3]{\ang*{#1}_{\! #2}^{\! #3}}    
\newcommand{\cjprod}[2]{\ang*{#1}^{\! #2}}           

\newcommand{\vo}[1]{\mathrm{V}_{\! #1}}         
\newcommand{\vop}[2]{\vo{#1}\brac*{#2}}         
\newcommand{\SCR}{\mathcal{S}}                  
\newcommand{\scr}[1]{\SCR_{#1}}                 
\newcommand{\scrs}[2]{\SCR_{#1}^{[#2]}}         
\newcommand{\cyc}[1]{\Gamma(#1)}                

\newcommand{\cfts}{conformal field theories}

\newcommand{\voa}{vertex operator algebra}
\newcommand{\voas}{vertex operator algebras}

\newcommand{\va}{vertex algebra}

\newcommand{\hw}{highest-weight}  
\newcommand{\hwv}{\hw{} vector}   
\newcommand{\hwvs}{\hw{} vectors} 

\newcommand{\sv}{singular vector}
\newcommand{\svs}{singular vectors}

\newcommand{\ope}{operator product expansion}

\newcommand{\rhs}{right-hand side}

\theoremstyle{plain}

\newtheorem*{thm*}{Theorem}

\Crefname{thm}{Theorem}{Theorems}
\Crefname{prop}{Proposition}{Propositions}
\Crefname{lem}{Lemma}{Lemmas}
\Crefname{cor}{Corollary}{Corollaries}
\Crefname{defn}{Definition}{Definitions}

\newcommand{\dprod}[4]{\prod_{#1}^{#2} \prod_{#3}^{#4}} 
\newcommand{\diff}{\dprod{k=1}{2}{i=1}{r_k} \dd z^k_i}  
\newcommand{\diffz}{\dprod{k=1}{2}{i=1}{r_k} \frac{\dd z^k_i}{z^k_i}} 
\newcommand{\diffn}{\dprod{k=1}{N-1}{i=1}{r_k} \dd z^k_i}
\newcommand{\diffnz}{\dprod{k=1}{N-1}{i=1}{r_k} \frac{\dd z^k_i}{z^k_i}}

\begin{document}

\title{Singular vectors for the \Walg{N} algebras}

\author[D Ridout]{David Ridout}

\address[David Ridout]{
School of Mathematics and Statistics \\
University of Melbourne \\
Parkville, Australia, 3010.
}

\email{david.ridout@unimelb.edu.au}

\author[S Siu]{Steve Siu}

\address[Steve Siu]{
School of Mathematics and Statistics \\
University of Melbourne \\
Parkville, Australia, 3010.
}

\email{ssiu1@student.unimelb.edu.au}

\author[S Wood]{Simon Wood}

\address[Simon Wood]{
School of Mathematics \\
Cardiff University \\
Cardiff, United Kingdom, CF24 4AG.
}

\email{woodsi@cardiff.ac.uk}

\thanks{\today}

\begin{abstract}
  In this paper, we use free field realisations of the A-type principal, or Casimir, \Walg{N} algebras to derive explicit formulae for singular vectors in Fock modules.  These singular vectors are constructed by applying screening operators to Fock module highest weight vectors. The action of the screening operators is then explicitly evaluated in terms of Jack symmetric functions and their skew analogues.  The resulting formulae depend on sequences of pairs of integers that completely determine the Fock module as well as the Jack symmetric functions.
\end{abstract}

\maketitle

\onehalfspacing

\section{Introduction} \label{sec:Intro}

A W-algebra is, generally speaking, a \voa{} whose set of generating fields includes conformal primaries of conformal weights greater than $1$. Aside from the Virasoro minimal models, such algebras were first considered by Zamolodchikov \cite{ZamInf85} shortly after the genesis of conformal field theory \cite{BPZCFT84}. We refer to \cite{BSWsym93} for an overview of W-algebras and their role in conformal field theory.

Here, we focus on the so-called \Walg{N} algebras, where \(N\ge 2\), which were first considered in \cite{BaiExt88}.  They form a distinguished family of W-algebras that are generated by \(N-1\) fields of conformal weights \(2,3,\dots, N\), where the generating field of conformal weight $2$ is the energy momentum tensor and all the remaining generators are conformal primaries.  While these algebras have received a lot of attention, there is much that remains poorly understood.  Even for the case $N=3$ there are significant difficulties, see \cite{BSW3} for example.

The \Walg{N} algebras were first proposed in the context of statistical mechanics, where they were linked to the continuum limits of certain \(\ZZ_N\)-symmetric lattice models \cite{FatZn85}.  In particular, $\wvoa$ was first introduced to describe the extended symmetry of the $3$ state Potts model \cite{ZamInf85}.  They enjoyed a period of intense popularity shortly thereafter as physicists explored the possibilities in their largely unsuccessful quest to classify all rational \cfts{}.  More recently, under the name of minimal model holography, \Walg{N} algebras (and their supersymmetric generalisations) have been the focus of intense scrutiny because of their role in constructing the AdS/CFT duals of Vasiliev higher spin theories of gravity on \(\grp{AdS}_3\) \cite{VasHS96}, see \cite{GabMMH13} for an in depth review.

The \Walg{N} algebras are also of great mathematical interest, since they are closely related to \voas{} constructed from affine Lie algebras through either cosets \cite{GKO85,BaiCos88} of the form
\begin{equation}
  \frac{\AKMA{sl}{N}_k\otimes\AKMA{sl}{N}_1}{\AKMA{sl}{N}_{k+1}}
\end{equation}
or through quantum hamiltonian reductions associated with principal nilpotents of \(\SLA{sl}{N}\) \cite{FeiQDS90,deBRel94}.  Although the equivalence of the coset and principal reduction pictures is only now being established rigorously \cite{AraCos17,ACL18}, this has the happy consequence that even though \Walg{N} algebras are quite different, structurally, to affine \voas{}, Lie-theoretic ideas can still be used to analyse them.

The purpose of this paper is to derive explicit formulae for singular vectors of the \Walg{N} algebras. Recall that the (universal) \Walg{N} algebra may be realised as a subalgebra of the rank \(N-1\) Heisenberg \va{} and thus, by restriction, the Fock spaces of the Heisenberg algebra become modules over the \Walg{N} algebras. It is within these Fock spaces that we derive explicit formulae for \Walg{N} singular vectors.

The derivation uses the simple, but far-reaching, fact that the universal enveloping algebra of the creation operators of a Heisenberg algebra is, as an associative algebra, isomorphic to the complexification of a ring of symmetric functions.\footnote{We adhere to the convention that a symmetric polynomial in a countably infinite number of variables is called a symmetric function.}  The appeal of this fact is that it allows one to express complicated singular vectors, consisting of large linear combinations of products of Heisenberg creation operators, as images of symmetric functions that have a very simple form under the aforementioned isomorphism.

This identification of singular vectors with symmetric functions can be traced back to Wakimoto and Yamada \cite{WakIrr83}, who discovered that Virasoro singular vectors in Fock spaces for the rank $1$ Heisenberg algebra of central charge \(c=1\) can be elegantly expressed in terms of Schur functions, a much studied basis of the ring of symmetric functions. This work was later generalised to free field realisations of the Virasoro algebra at arbitrary central charge by Mimachi and Yamada \cite{MimJa95}, with Schur functions being replaced by Jack symmetric functions \cite{JacCla70}.  We recall that the Jack functions form a one-parameter family of bases of the ring of symmetric functions and that the Schur functions correspond to the parameter being set to $1$.

This work relied heavily on a construction by Tsuchiya and Kanie \cite{TKScr286,TKScr86} of Virasoro module homomorphisms called screening operators.  With these, singular vectors may be realised as images of Fock space \hwvs{}.  These screening operators are also the key ingredient in Dotsenko and Fateev's Coulomb gas formalism \cite{DotScr84}.  Similar \sv{} constructions using screening operators have recently been detailed for the free field realisations of the \voas{} associated with \(\AKMA{sl}{2}\) and the \(\mathcal{N}=1\) superconformal algebras, with singular vectors being evaluated using Jack symmetric functions, their supersymmetric generalisations and related families \cite{KatMis92,DesSup01,DesSJa12,RWsl215,YanUgl15,OPDS16}.

The Jack symmetric function basis is indexed by partitions of integers and, interestingly, the singular vectors of Fock spaces for the Virasoro algebra and \(\AKMA{sl}{2}\) are always associated to a single Jack symmetric function indexed by a rectangular partition, that is, a partition whose parts are all equal. It is therefore natural to try and find an interpretation for the Jack symmetric functions indexed by non-rectangular partitions. One such interpretation was found by Awata, Odake, Matsuo and Shiraishi \cite{AwaCS95} while studying a connection between the Calogero-Sutherland model, a model for a system of non-relativistic particles on a circle with an inverse square potential, and the \Walg{N} algebras.  In \cite{AwaCS95}, it is noted that \Walg{N} \svs{} can be realised using screening operators acting on rank $N-1$ Fock spaces.  However, these singular vectors were not explicitly evaluated, but were instead projected onto a rank $1$ Fock space, where they were identified as eigenstates of the hamiltonian of the Calogero-Sutherland model.  As the eigenstates of the Calogero-Sutherland model may be expressed as products of a ground state wavefunction and Jack polynomials, the conclusion of \cite{AwaCS95} is then that \Walg{N} \svs{} can be used to construct arbitrary Jack polynomials.

Here, in contrast to \cite{AwaCS95}, we do not project, but stay in the rank \(N-1\) Fock spaces, using symmetric function theory to find explicit formulae for \Walg{N} singular vectors.  Our main result is that certain singular vectors in rank $N-1$ Fock spaces may be identified with linear combinations of Jack symmetric functions and their skew variants, parametrised by sequences of partitions.

The motivation for this study of singular vectors is that they have proven invaluable in finding short elegant proofs of the classification theorems for modules of rank $1$ \voas{} \cite{TsuExt13,RWVir15,RWsl215,BloN117}.  We believe that a solid understanding of higher rank singular vectors will pave the way to similar classification theorems for higher rank \voas{}.

This paper is organised as follows. \cref{sec:heisva,sec:W3,sec:symmfns} form an overview of known results and serve to fix notation.  In \cref{sec:heisva}, we review the rank \(r\) Heisenberg vertex algebra for later use.  We show how energy-momentum tensors can be constructed for arbitrary choices of central charge and discuss vertex operators and their compositions.  In \cref{sec:W3}, we give the standard free field realisation of the \Walg{3} algebras by explicitly constructing them as subalgebras of the rank $2$ Heisenberg vertex algebra.  We then determine their screening operators and identify the Fock spaces on which we can consistently evaluate their action.  \cref{sec:symmfns} gives a brief overview of the ring of symmetric functions, with particular emphasis on the Jack symmetric function bases, these being the bases required to evaluate the action of screening operators.

The main results, explicit formulae for \Walg{N} singular vectors in certain Fock spaces in terms of the Jack symmetric functions, are given in \cref{sec:W3eval,sec:examples,sec:WN} --- see \cref{eq:W3SVFormula} for the \Walg{3} singular vector formula and Equation \eqref{eq:WNSVFormula} for that of \Walg{N}. In \cref{sec:W3eval}, we give a detailed derivation of these formulae for the \Walg{3} algebras.  This case already illustrates the complexity of the general computations while keeping the formulae reasonably brief.  This is followed by a discussion of simple examples in \cref{sec:examples} featuring low-grade \svs{}.  Although we do not report the results of investigating any really complicated examples, symbolic algebra packages incorporating symmetric functions, \textsc{SageMath} for example, allow such investigations to be straightforwardly performed to quite high grades.  In \cref{sec:WN}, we generalise the singular vector formulae of \cref{sec:W3eval} to the \Walg{N} algebras for all \(N\ge3\).

\section{The rank $r$ Heisenberg algebra}
\label{sec:heisva}

The theory of \(r\) free chiral bosons, also known as the rank \(r\) Heisenberg vertex algebra \(\hvoa{r}\), is a staple of conformal field theory and \voa{} theory. Here, we construct the Heisenberg algebras by affinising abelian Lie algebras.

The rank $r$ Heisenberg algebra is
constructed from an \(r\)-dimensional complex vector space \(\hvs\) together with a non-degenerate symmetric bilinear form \(\kil{\blank}{\blank}\). We pick a basis \(\{\scrt{1},\dots,\scrt{r}\}\) of
\(\hvs\) such that the Gram matrix of \(\kil{\blank}{\blank}\) is the Cartan matrix of \(\SLA{sl}{r+1}\):
\begin{equation}
	\kil{\scrt{i}}{a^j}=2\delta_{i,j}-\delta_{i+1,j}-\delta_{i,j+1},\quad i,j=1,\dots,r.
\end{equation}
Since \(\kil{\blank}{\blank}\) is non-degenerate, it defines a vector space
isomorphism \(\iota:\hvs\to\dvs\) by \(a\mapsto\kil{a}{\blank}\).
The induced non-degenerate symmetric bilinear form will also be denoted by \(\kil{\blank}{\blank}\). We denote the
images of the basis vectors \(\scrt{i}\) by \(\srt{i}=\iota(\scrt{i})\) and the elements of the basis of \(\dvs\) dual to \(\set{\scrt{i}}\) by \(\fwt{i}\).  Thus, $\fwt{i}(\scrt{j}) = \delta_i^j$.
The \(\srt{i}\) and \(\fwt{i}\) may therefore be identified as simple roots and fundamental weights, respectively, of \(\SLA{sl}{r+1}\). In this picture, the basis vectors \(\scrt{i}\in\hvs\) are the simple coroots of \(\SLA{sl}{r+1}\).

To any vector \(a\in\hvs\), one assigns a field \(a(z)\) whose defining operator product expansions are
\begin{equation}
	a(z)b(w)\sim \frac{\kil{a}{b}}{(z-w)^2}, \quad a,b\in\hvs.
\end{equation}
These fields admit Fourier expansions of the form
\begin{equation}
	a(z)=\sum_{n\in\ZZ} a_{n} z^{-n-1},\quad a\in\hvs,
\end{equation}
whose modes satisfy the following commutation relations:
\begin{equation}
	\comm[\big]{a_m}{b_n}=m\kil{a}{b} \delta_{m,-n}\wun.
\end{equation}
The Heisenberg Lie algebra \(\heis\) is the infinite-dimensional Lie algebra spanned
by the central element \(\wun\) and the generators \(a_m\), for all
\(a\in\hvs\) and \(m\in\ZZ\). We have chosen to denote the central
element by \(\wun\), since we assume that it will act as the identity on any \(\heis\)-module.
A basis of $\heis$ is then given by $\wun$ and the $\scrt{i}_m$, with $i=1,\dots,r$ and $m\in\ZZ$.

The Heisenberg Lie algebra admits a triangular decomposition
\begin{align}
	\heis=\heis_{-}\oplus \heis_0\oplus\heis_+,
	\quad \heis_\pm=\bigoplus_{i=1}^r\bigoplus_{m\ge 1}\CC \scrt{i}_{\pm m},
	\quad \heis_0=\bigoplus_{i=1}^r\CC \scrt{i}_0\oplus\CC\wun.
\end{align}
Verma modules over $\heis$ are commonly referred to as Fock spaces. These are induced from the one-dimensional modules $\CC\ket{\zeta}$, $\zeta\in\dvs$, over $\heis_{\ge}=\heis_0\oplus\heis_+$ that are defined by
\begin{align}
	\wun\ket{\zeta}=\ket{\zeta}, \qquad
  a_n\ket{\zeta}=\delta_{n,0} \zeta(a) \ket{\zeta},\quad a\in\hvs,\quad n\ge0.
\end{align}
The Fock spaces
\begin{align}
  \Fock{\zeta}=\UEA{\heis}\otimes_{\UEA{\heis_\ge}}\CC\ket{\zeta}
\end{align}
are well known to be simple $\heis$-modules, for all $\zeta \in \dvs$.

As a module over itself, the Heisenberg vertex algebra $\hvoa{r}$ is identified with
the Fock space \(\Fock{0}\) and the state-field correspondence is given by
\begin{equation}
	\ket{0}\longleftrightarrow \wun,
	\quad b^1_{-n_1-1}\cdots b^k_{-n_k-1}\ket{0}\longleftrightarrow
	\normord{\frac{\partial^{n_1}}{n_1!}b^1(z)\cdots\frac{\partial^{n_k}}{n_k!}b^k(z)},
\end{equation}
where \(b^1,\dots,b^k\in \hvs\) and normal ordering is defined in the usual way.

The Heisenberg vertex algebra \(\hvoa{r}\) can be endowed with the structure
of a \voa{} by choosing an energy-momentum tensor. This choice is not
unique. For the purposes of this note, we shall restrict our attention to the following one-parameter family of energy-momentum tensors:
\begin{equation}\label{eq:emtensor}
	T(z) = \sum_{i=1}^r \sqbrac*{\frac{1}{2} \normord{\scrt{i}(z) \dscrt{i}(z)} +
		\alpha_0 \partial \dscrt{i}(z)}, \quad \alpha_0 \in \CC.
\end{equation}
Here, the \(\dscrt{i} \in \hvs\) are dual to the coroots \(\scrt{i}\) in the sense that $\iota(\dscrt{i})=\fwt{i}$.
We note that while the quadratic summand in the above energy-momentum tensor
is basis independent, the linear summand is not. The central charge
corresponding to this choice of energy-momentum tensor depends on the
parameter \(\alpha_0\):
\begin{equation}\label{eq:ccharge}
	c = r - r(r+1)(r+2) \alpha_0^2.
\end{equation}
By definition, the coefficients of the Fourier expansion of the
energy-momentum tensor satisfy the commutation relations of the Virasoro
algebra. Thus, formula \eqref{eq:emtensor} realises the Virasoro generators
\(L_n\), $n \in \ZZ$, as infinite sums of products of Heisenberg generators:
\begin{align}
	T(z)=\sum_{n\in \ZZ}L_n z^{-n-2}, \quad
	L_n=\sum_{i=1}^r\sqbrac*{\frac{1}{2}\sum_{m\in\ZZ}
		\normord{\scrt{i}_m\dscrt{i}_{n-m}}-\alpha_0(n+1)\dscrt{i}_n}.
\end{align}
This identification yields an action of the Virasoro algebra on the Fock
spaces \(\Fock{\zeta}\), \(\zeta\in\dvs\). In this way, any \hwv{} \(\ket{\zeta} \in \Fock{\zeta}\) is also a Virasoro \hwv{}:
\begin{align} \label{eq:DefConfWt}
	L_n\ket{\zeta}=h_\zeta\delta_{n,0}\ket{\zeta},\quad n\ge 0,\quad
	h_\zeta=\frac{1}{2} \kil{\zeta}{\zeta-2\alpha_0\weyl}.
\end{align}
Here, $\weyl = \sum_i\fwt{i}$ is the Weyl vector of $\SLA{sl}{r+1}$.  We note that while the Fock spaces are simple as Heisenberg modules, they need not be as Virasoro modules.

The primary fields of the free boson theory are called vertex operators (not to be confused with elements of the
Heisenberg vertex operator algebra). To define them, we first need to extend $\heis$ by \(\CC[\dvs]\), the group algebra of \(\dvs\), treating \(\dvs\) as an abelian group under vector addition and $\CC[\dvs]$ as an abelian Lie algebra.
We denote the group algebra basis element corresponding to \(\eta\in\dvs\) by \(\gae{\eta}\) and define the commutation relations between the generators \(a_m\) and \(\gae{\eta}\) by
\begin{align}
	\comm*{a_m}{\gae{\eta}}=\delta_{m,0} \eta(a)\gae{\eta}, \quad a\in\hvs, \quad \eta\in\dvs, \quad m\in\ZZ.
\end{align}
It is easy to check that this extension of $\heis$ by $\CC[\dvs]$ is a semidirect sum of Lie algebras.

A standard computation now shows that $\gae{\eta}$ maps the \hwv{}
$\ket{\zeta} \in \Fock{\zeta}$ to a \hwv{} of $a_0$-eigenvalue
$\zeta(a) + \eta(a)=(\zeta+\eta)(a)$.
Following usual practice, we shall
identify $\gae{\eta} \ket{\zeta}$ with $\ket{\zeta + \eta}$.
The vertex operator corresponding to $\ket{\zeta} = \gae{\zeta} \ket{0}$ is
\begin{align} \label{eq:DefVOps}
\vop{\zeta}{z}=\gae{\zeta}z^{a_0}\prod_{m\ge1}\exp\brac*{\frac{a_{-m}}{m}z^{m}} \exp\brac*{-\frac{a_m}{m}z^{-m}}, \quad \zeta = \iota(a) \in \dvs.
\end{align}
These primary fields therefore define linear maps between Fock spaces:
\begin{align}
  \vop{\zeta}{z}\colon\Fock{\eta}\to z^{(\zeta,\eta)} \Fock{\zeta+\eta}\llbracket z,z^{-1}\rrbracket.
\end{align}
It is easy to check from the $\hvoa{r}$-primary \ope{}
\begin{align}\label{ope:av}
  a(z)\vop{\zeta}{w}&\sim \frac{\zeta(a)\vop{\zeta}{w}}{z-w}
\end{align}
that $a(z)$ and $\vop{\zeta}{w}$ are mutually local for all $a \in \hvs$ and $\zeta \in \dvs$.  The same is therefore true for an arbitrary field of $\hvoa{r}$ and any vertex operator, by Dong's lemma.

Finally, suppose that $\zeta_i = \iota(b^i) \in \dvs$, for $i=1,\dots,k$.
Then, a standard computation allows one to write the composition of the $k$ vertex operators $\vop{\zeta_i}{z_i}$ as
\begin{align}\label{eq:vops}
\vop{\zeta_1}{z_1}\cdots \vop{\zeta_k}{z_k}=\prod_{i=1}^k
\gae{\zeta_i} \cdot \prod_{\mathclap{1\le i<j\le k}} \:
(z_i-z_j)^{(\zeta_i,\zeta_j)} \cdot \prod^k_{i=1} z_i^{b^i_0} \cdot
\prod_{m\ge 1}\exp\brac*{\frac{1}{m}\sum^k_{i=1}b^i_{-m}z^m_i} \exp\brac*{-\frac{1}{m}\sum^k_{i=1}b^i_{m}z^{-m}_i}.
\end{align}
This explicit formula will be used many times in what follows.

\section{The Universal \Walg{3} Vertex Operator Algebra}
\label{sec:W3}

In this section, we restrict ourselves to the rank $2$ Heisenberg vertex algebra \(\hvoa{2}\) and, in the vein of \cite{ZamZ85}, define a family of subalgebras, each denoted by \(\Wvoa{3}\), called the \Walg{3} vertex operator algebras, or \Walg{3} algebras for short. These algebras are parametrised by $\alpha_0 \in \CC$ and are strongly generated by the energy-momentum tensor defined in \eqref{eq:emtensor} and an additional primary field \(W(z)\) of conformal weight $3$.

In the basis \(\set{\scrt{1},\scrt{2}}\) defined above, for which the Gram matrix of
the inner product \(\kil{\blank}{\blank}\) is the Cartan matrix of
\(\SLA{sl}{3}\), the energy-momentum tensor \(T(z)\) is
\begin{subequations} \label{eq:ffr}
	\begin{align}
		T(z)&=\frac{1}{3}\normord{\scrt{1}(z)\scrt{1}(z)}+\frac{1}{3}\normord{\scrt{1}(z)\scrt{2}(z)}+\frac{1}{3}\normord{\scrt{2}(z)\scrt{2}(z)}
		+\alpha_0\partial_z \scrt{1}(z)+\alpha_0\partial_z \scrt{2}(z)
	\intertext{and the central charge is $c=2-24\alpha_0^2$.  The conformal primary of weight $3$ is then}
		W(z)&=\frac{\sqrt{\beta}}{18\sqrt{3}}\Bigl[
		 2\normord{\brac*{\scrt{2}(z)-\scrt{1}(z)}\brac*{\scrt{1}(z)+2\scrt{2}(z)}\brac*{2\scrt{1}(z)+\scrt{2}(z)}} \Bigr.\nonumber\\
		 &\quad\Bigl. +9\alpha_0\brac*{\normord{\partial \scrt{2}(z)\brac*{\scrt{1}(z)+2\scrt{2}(z)}}
			-\normord{\partial \scrt{1}(z)\brac*{2\scrt{1}(z)+\scrt{2}(z)}}}
		+9\alpha_0^2\brac*{\partial^2 \scrt{2}(z)-\partial^2\scrt{1}(z)}\Bigr],
	\end{align}
\end{subequations}
where
\begin{equation}
	\beta=\frac{16}{22+5c} = \frac{2}{4-15\alpha_0^2}
\end{equation}
in the conventional normalisation, appropriate for $c \neq -\frac{22}{5}$ ($\alpha_0 \neq \pm \frac{2}{\sqrt{15}}$). A somewhat involved computation now determines the operator product expansion of \(W(z)\) with itself to be
\begin{align}
	W(z)W(w)&\sim\frac{c/3}{(z-w)^6}+\frac{2T(w)}{(z-w)^4}+\frac{\partial T(w)}{(z-w)^3}+\frac{\frac{3}{10}\partial^2 T(w)+2\beta\Lambda(w)}{(z-w)^2}+\frac{\frac{1}{15}\partial^3T(w)+\beta\partial\Lambda(w)}{z-w},
\end{align}
where \(\Lambda(z)=\normord{T(z)T(z)}-\frac{3}{10}\partial^2 T(z)\).  This, along with the primary nature of $W(z)$, implies the commutation relations
\begin{subequations} \label{eq:W3ModeAlgebra}
	\begin{align}
		\comm[\big]{L_m}{W_n}&=(2m-n)W_{m+n},\\
		\comm[\big]{W_m}{W_n}&=(m-n)\sqbrac*{\frac{1}{15}(m+n+3)(m+n+2)-\frac{1}{6}(m+2)(n+2)}L_{m+n}\nonumber\\
		&\quad+\beta(m-n)\Lambda_{m+n}+\frac{c}{360}m(m^2-1)(m^2-4)\delta_{m+n,0},
	\end{align}
\end{subequations}
where $W(z) = \sum_{n \in \ZZ} W_n z^{-n-3}$.

Since Fock spaces are modules over the Heisenberg \voa{} \(\hvoa{2}\) and
we have defined the \Walg{3} algebra as a subalgebra of \(\hvoa{2}\), each Fock space is a \(\Wvoa{3}\)-module, by restriction. In particular, the \hwv{} \(\ket{\zeta}\in\Fock{\zeta}\),
\(\zeta\in\dvs\), is also a \hwv{} for \(\Wvoa{3}\):
\begin{align}
  L_n\ket{\zeta}= \delta_{n,0}h_\zeta \ket{\zeta},\quad
  W_n\ket{\zeta}= \delta_{n,0}w_\zeta \ket{\zeta},\quad n\ge0.
\end{align}
Here, $h_\zeta$ was given in \eqref{eq:DefConfWt} and the $W_0$-eigenvalue is given by
\begin{align} \label{eq:DefWWt}
  w_\zeta &= \sqrt{3\beta} \kil{\zeta}{\fwt{2}-\fwt{1}} \brac[\big]{\kil{\zeta}{\fwt{1}} - \alpha_0} \brac[\big]{\kil{\zeta}{\fwt{2}} - \alpha_0}.
\end{align}

Our main reason for introducing vertex operators in \eqref{eq:DefVOps} is to construct linear maps between Fock spaces that commute with the
action of an appropriate subalgebra of the Heisenberg vertex algebra. Here, we wish to
construct maps that commute with \(\Wvoa{3}\), that is, \(\Wvoa{3}\)-module homomorphisms.  Such module homomorphisms are called \emph{screening operators} and they are constructed from \emph{screening fields}, these being
vertex operators whose operator product expansions with the fields of \(\Wvoa{3}\) are total derivatives. For this, it clearly suffices to find fields whose operator product expansions with the generating fields \(T(z)\) and \(W(z)\) are total derivatives.

As the vertex operator $\vop{\zeta}{w}$ is a conformal primary of weight $h_{\zeta}$, its \ope{} with $T(z)$ will be a total derivative if and only if \(h_{\zeta}=1\). Unsurprisingly, the analogous computation for \(W(z)\) is more involved (we used Thielemans' \textsc{OPEdefs} package for \textsc{Mathematica}.
We shall not give the unpleasant details, noting instead that a necessary condition for the operator product expansion \(W(z) \vop{\zeta}{w}\) to be a total derivative is that the coefficient of $(z-w)^{-1}$ in this expansion is a total derivative.
Analysing this explicitly, for general \(\zeta\in\dvs\), and recalling that $h_{\zeta} = 1$, we conclude that this coefficient will be a total derivative if \(\zeta_1=0\), \(\zeta_2=0\) or if \(\zeta_1=\zeta_2\) and \(\alpha_0=0\) (where the $\zeta_i$ denote Dynkin labels:  $\zeta = \sum_i \zeta_i \fwt{i}$). As we are only interested in screening operators that exist for all values of \(\alpha_0\), it follows that there are exactly four possible weights $\zeta$ that can be used to construct screening operators: $\zeta = \alpha_{\pm} \srt{1}, \alpha_{\pm} \srt{2}$.
Here, we define
\begin{align}\label{eq:alphadef}
	\alpha_+=\frac{1}{2} \brac*{\alpha_0+\sqrt{\alpha_0^2+4}},\quad
	\alpha_-=\frac{1}{2} \brac*{\alpha_0-\sqrt{\alpha_0^2+4}}
\end{align}
to be the solutions of the quadratic equations \(h_{\zeta_i\srt{i}}=1\), for \(i=1,2\).
It only remains to confirm that the full operator product expansion \(W(z) \vop{\zeta}{w}\), when $\zeta$ is one of the above weights, is indeed a total derivative:
\begin{align}
	W(z)\vop{\alpha_\pm\srt{i}}{w}\sim-(-1)^i\sqrt{\frac{6}{4-15\alpha_0^2}}\partial_w
	\brac*{
		\frac{\alpha_0\vop{\alpha_\pm\srt{i}}{w}}{2(z-w)^2}-\frac{\normord{\dscrt{1}(w)\vop{\alpha_\pm\srt{i}}{w}}}{z-w}
		},
		\quad i=1,2.
\end{align}

Having identified screening fields for \(\Wvoa{3}\), we construct screening operators by taking residues:
\begin{equation}
	\scr{\pm i}=\oint_0 \vop{\alpha_\pm\srt{i}}{w} \,\dd w.
\end{equation}
Here, the residue is indicated using a simple anticlockwise contour that encircles $0$ once (we absorb the usual factor of $2\pi\ii$ into the definition of the contour integral).
These screening operators define \(\Wvoa{3}\)-module homomorphisms since
\begin{align}
	\comm[\big]{T(z)}{\scr{\pm i}}
	=-\oint_z T(z)\vop{\alpha_\pm \srt{i}}{w} \,\dd w=0, \quad 	\comm[\big]{W(z)}{\scr{\pm i}}
	=-\oint_z W(z)\vop{\alpha_\pm \srt{i}}{w} \,\dd w=0.
\end{align}
These identities follow from the mutual locality of Heisenberg fields and vertex operators, see \eqref{ope:av}.

Taking the residue of a screening field \(\vop{\alpha_\pm\srt{i}}{z}\) is
of course only well defined when it is acting on a $\hvoa{2}$-module for which the exponents of \(z\) in the Fourier expansion of \(\vop{\alpha_\pm\srt{i}}{z}\) are all integers. In case the $\hvoa{2}$-module is the Fock space $\Fock{\eta}$, this is satisfied if and only if \(\alpha_{\pm}\kil{\srt{i}}{\eta}\in\ZZ\). Fortunately, one can also construct screening operators by integrating compositions \eqref{eq:vops}
of multiple screening fields. In particular, composing \(r_2\) copies of \(\vop{\alpha_\pm \srt{2}}{w}\) with \(r_1\) copies of \(\vop{\alpha_\pm\srt{1}}{z}\) and then acting on $\Fock{\eta}$ gives
\begin{multline}\label{eq:scrfields}
	\Bigl.\vop{\alpha_\pm\srt{1}}{z_1}\cdots\vop{\alpha_\pm\srt{1}}{z_{r_1}}
	\vop{\alpha_\pm \srt{2}}{w_1}\cdots\vop{\alpha_\pm\srt{2}}{w_{r_2}}\Bigr\rvert_{\Fock{\eta}}\\
	=\prod_{1\le i<j\le r_1}\brac*{z_i-z_j}^{2\alpha_\pm^2}
	\cdot\prod_{1\le i<j\le r_2}\brac*{w_i-w_j}^{2\alpha_\pm^2}
	\cdot\prod_{i=1}^{r_1}\prod_{j=1}^{r_2}\brac*{z_i-w_j}^{-\alpha\pm^2}
	\cdot\prod_{i=1}^{r_1}z_i^{\alpha_\pm\kil{\srt{1}}{\eta}}
	\cdot\prod_{j=1}^{r_2}w_j^{\alpha_\pm\kil{\srt{2}}{\eta}}\\
	\quad\cdot
	\gae{r_1\alpha_\pm\srt{1}+r_2\alpha_\pm\srt{2}} \prod_{m\ge1}
	\exp\sqbrac*{\alpha_\pm\brac*{
			\scrt{1}_{-m}\sum_{i=1}^{r_1}\frac{z_i^m}{m}
			+\scrt{2}_{-m}\sum_{i=1}^{r_2}\frac{w_i^m}{m}
		}}
	\exp\sqbrac*{-\alpha_\pm\brac*{
			\scrt{1}_{m}\sum_{i=1}^{r_1}\frac{z_i^{-m}}{m}
			+\scrt{2}_{m}\sum_{i=1}^{r_2}\frac{w_i^{-m}}{m}
		}}.
\end{multline}
Up to a complex phase, which we suppress, the first five multivalued factors in this expression can be rewritten in the form
\begin{multline}\label{eq:mvfn}
	\prod_{1\le i\neq j\le r_1}\brac*{1-\frac{z_i}{z_j}}^{\alpha_\pm^2}
	\cdot\prod_{1\le i\neq j\le r_2}\brac*{1-\frac{w_i}{w_j}}^{\alpha_\pm^2}
	\cdot\prod_{i=1}^{r_1}\prod_{j=1}^{r_2}\brac*{1-\frac{w_j}{z_i}}^{-\alpha_\pm^2}\\
	\cdot\prod_{i=1}^{r_1}z_i^{\alpha_\pm\kil{\srt{1}}{\eta}+\alpha_\pm^2\brac*{r_1-r_2-1}}
	\cdot\prod_{j=1}^{r_2}w_j^{\alpha_\pm\kil{\srt{2}}{\eta}+\alpha_\pm^2\brac*{r_2-1}},
\end{multline}
thereby isolating the non-integer exponents of the $z_i$ and $w_j$ in the last two factors.

Finding closed (multivariable) contours over which multivalued functions such as
\eqref{eq:mvfn} can be integrated (to obtain $\Wvoa{3}$-module homomorphisms) is a highly non-trivial problem.
Fortunately, Tsuchiya and Kanie solved this problem for the rank $1$
Heisenberg vertex algebra \cite{TKScr86} by constructing cycles with non-trivial homology classes over which screening operators can be integrated.
These cycles, which we shall denote by \(\cyc{m;t}\) for \(m\in \ZZ_{\ge0}\)
and \(t\in\CC\setminus \QQ_{\le0}\),\footnote{
	The range of the
  parameter \(t\) could in principle be extended to \(\CC\setminus\{0\}\).
  However, to avoid singularities in certain coefficients, this would require
  one to use a different normalisation of the Jack symmetric function basis
  presented in \cref{sec:symmfns}. Moreover,
  some linear independence arguments would become
  more complicated. For simplicity, we therefore
  avoid non-positive rational values of the parameter \(t\).}
allow one to integrate expressions of the form
\begin{equation}
	\int_{\cyc{m;t}}\prod_{1\le i\neq j\le m}\brac*{1-\frac{z_i}{z_j}}^{1/t}
	\cdot f(z) \,\dd z_1\cdots \dd z_m,
\end{equation}
where \(f(z)\) is a Laurent polynomial in \(z_1,\dots,z_m\) which is invariant with respect to permuting the indices of its variables. We shall not describe the construction of these cycles in any detail.  It will, however, be convenient to normalise them by requiring that
\begin{equation}\label{eq:cycnorm}
	\int_{\cyc{m;t}}\prod_{1\le i\neq j\le m}\brac*{1-\frac{z_i}{z_j}}^{1/t}
	 \frac{\dd z_1\cdots \dd z_m}{z_1 \cdots z_m}=1.
\end{equation}

The cycles \(\cyc{m;t}\) can be used to construct screening operators from the compositions \eqref{eq:scrfields} whenever the exponents of the \(z_i\) and \(w_j\) are integers. If this is the case, then the screening operators are defined as
\begin{equation}\label{eq:scrs}
	\scrs{\pm}{r_1,r_2}=\int_{\cyc{r_1;1/\alpha_\pm^2}}\int_{\cyc{r_2;1/\alpha_\pm^2}}
	\vop{\alpha_\pm\srt{1}}{z_1}\cdots\vop{\alpha_\pm\srt{1}}{z_{r_1}}
	\vop{\alpha_\pm \srt{2}}{w_1}\cdots\vop{\alpha_\pm
		\srt{2}}{w_{r_2}}
	\,\dd z_1\cdots \dd z_{r_1}\dd w_1\cdots\dd w_{r_2}.
\end{equation}
By construction, these screening operators are \(\Wvoa{3}\)-module
homomorphisms when acting on appropriate Fock spaces.

We parametrise the Fock space weights for which the screening operators \eqref{eq:scrs} are defined as follows:
\begin{equation} \label{eq:defsvwts}
	\svwts{u_1,v_1}{u_2,v_2}=\brac[\big]{\brac{1-u_1}\alpha_+	+ \brac{1-v_1}\alpha_-}\fwt{1}
	+\brac[\big]{\brac{1-u_2}\alpha_+ + \brac{1-v_2}\alpha_-}\fwt{2}, \quad u_1,u_2,v_1,v_2\in\ZZ.
\end{equation}
Considering the exponents of the last two factors of \eqref{eq:mvfn}, we conclude that the screening operators define $\Wvoa{3}$-module homomorphisms between the following Fock spaces:
\begin{equation} \label{eq:welldefined}
	\begin{aligned}
		\scrs{+}{r_1,r_2}\colon\Fock{\svwts{r_1-r_2,s_1}{r_2,s_2}}&\to\Fock{\svwts{-r_1,s_1}{r_1-r_2,s_2}}, &
		r_1,r_2&\in\ZZ_{\ge0}, & s_1,s_2&\in\ZZ,\\
		\scrs{-}{s_1,s_2}\colon\Fock{\svwts{r_1,s_1-s_2}{r_2,s_2}}&\to\Fock{\svwts{r_1,-s_1}{r_2,s_1-s_2}}, &
		r_1,r_2&\in\ZZ, & s_1,s_2&\in\ZZ_{\ge0}.
	\end{aligned}
\end{equation}
Evaluating the action of these screening operators initially appears rather daunting. However, we know from \eqref{eq:scrfields} that compositions of screening fields factorise into a product of a multivalued function and certain
power series in the \(z_i\) and \(w_j\) that are \emph{symmetric} with respect to permuting the \(z_i\) among themselves and, separately, the \(w_j\) among themselves. The theory of symmetric functions provides the tools that allow us to evaluate the action of these screening operators on certain Fock spaces.  We therefore turn to a discussion of these tools.

\section{The ring of symmetric functions}
\label{sec:symmfns}

The purpose of this section is to review various results from the theory of symmetric functions that will be used to evaluate the action of screening operators on certain Fock spaces.
The standard reference for symmetric functions and their myriad properties is Macdonald's book \cite{MacSym95} to which we refer the reader for more details.

Let $\fsym{n}$ denote the ring of symmetric polynomials in the \(n\) variables $z_1, \dots, z_n$. This is the subring of \(\CC[z_1,...,z_n]\) that consists of the polynomials that are invariant with respect to permuting the indices of the $z_i$. It admits numerous interesting generators such as the power sums
\begin{align}
  \powsum{k}=\sum_{i=1}^n z^k_i, \quad k\ge1.
\end{align}
For \(1\le k\le n\), the \(\powsum{k}\) are algebraically
independent and freely generate \(\fsym{n}\), that is,
\begin{align}
\fsym{n}=\CC[\powsum{1},\cdots,\powsum{n}].
\end{align}
We can therefore use partitions \(\lambda=\partn{\lambda_1,\lambda_2,\dots}\), whose
parts $\lambda_i$ are bounded by \(n\), to define
\begin{equation}
  \powsum{\lambda}=\powsum{\lambda_1}\cdots\powsum{\lambda_k}.
\end{equation}
These power sums, labelled by partitions whose parts do not exceed $n$, thus form a basis of
\(\fsym{n}\):
\begin{equation}\label{eq:powbasis}
  \fsym{n}=\bigoplus_{\lambda, \lambda_1\le n}\CC \powsum{\lambda}.
\end{equation}

Another family of symmetric polynomials is given by the monomial symmetric polynomials
\begin{equation}
  \monsym{\lambda}=\sum_{\sigma}z_1^{\lambda_{\sigma(1)}}\cdots z_n^{\lambda_{\sigma(n)}},
\end{equation}
where \(\sigma\) runs over all distinct permutations of the partition \(\lambda\).  In this case, \(\lambda\) is not constrained by a bound on its individual parts, but by their number \(\len{\lambda}\) (the length of \(\lambda\)) which is at most $n$. Note that each monomial summand of $\monsym{\lambda}$ has coefficient $1$. For example,
\begin{equation}
  \fmonsym{[2,2]}{z_1,z_2}=z_1^2z_2^2,\quad
  \fmonsym{[2,2]}{z_1,z_2,z_3}=z_1^2z_2^2+z_1^2z_3^2+z_2^2z_3^2.
\end{equation}
The monomial symmetric polynomials also form a basis of \(\fsym{n}\):
\begin{equation}\label{eq:monbasis}
  \fsym{n}=\bigoplus_{\lambda,\len{\lambda}\le n}\CC\monsym{\lambda}.
\end{equation}

The respective restrictions on parts and lengths of partitions in the definitions of these symmetric polynomials can be avoided by taking a formal limit to infinitely many variables. The
resulting ring \(\sym\) is called the ring of symmetric functions and, unsurprisingly, its
elements are called symmetric functions. The ring \(\fsym{n}\) of symmetric
polynomials in \(n\) variables can then be easily recovered from \(\sym\) by
setting all but the first \(n\) variables to $0$.  This amounts to a projection
\begin{align}
  \pi_n:\sym\to\fsym{n}, \quad
	f(x_1,x_2,\dots)\mapsto f(x_1,\dots,x_n,0,0,\dots).
\end{align}
In \(\sym\), the power sums \(\powsum{k}\) are algebraically
independent for all \(k\ge1\) and they freely generate \(\sym\), that is,
\begin{equation} \label{eq:sym=polyring}
  \sym=\CC[\powsum{1},\powsum{2},\dots].
\end{equation}
Similarly, the restrictions on the sizes of the parts and the lengths of the partitions labelling power sums and monomial
symmetric functions, respectively, no longer apply. Both classes of symmetric functions give bases of \(\sym\):
\begin{equation}
  \sym = \bigoplus_{\lambda}\CC \powsum{\lambda}=\bigoplus_{\lambda}\CC\monsym{\lambda}.
\end{equation}
We note that \(\pi_n\brac*{\monsym{\lambda}}=0\) if and only if \(\len{\lambda}>n\), but that no such truncations exist for the power sums \(\powsum{k}\): their images under \(\pi_n\) are all non-zero.

There exists another family of bases of \(\sym\) and \(\fsym{n}\) labelled by partitions, called the
Jack symmetric functions and Jack symmetric polynomials (or just Jack functions or
polynomials for short), respectively.  These are defined using the dominance partial ordering of partitions: if $\lambda$ and $\mu$ are both partitions of the same non-negative integer, then we write \(\lambda\ge \mu\) (and say that $\lambda$ dominates $\mu$) if
\begin{equation}
  \lambda_1+\cdots+\lambda_i\ge\mu_1+\cdots+\mu_i,
\end{equation}
for all \(i\ge1\).

For each $t\in\CC\setminus\QQ_{\le0}$ (the non-positive rationals are excluded to avoid certain normalisation problems),
the Jack functions \(\jack{\lambda}{t}\) are uniquely defined by the following two properties:
\begin{enumerate}
	\item For any partition \(\lambda\), \(\jack{\lambda}{t}\) admits an upper triangular decomposition of the form
  \begin{align}\label{upptri}
    \jack{\lambda}{t}=\monsym{\lambda}+\sum_{\lambda>\mu} v_{\lambda,\mu}(t)\monsym{\mu}, \qquad v_{\lambda,\mu}(t) \in \CC.
  \end{align}
	\item The Jack functions form an orthogonal basis of $\sym$ with respect to the inner product defined by
  \begin{align}\label{innprod}
    \cjprod{\powsum{\lambda},\powsum{\mu}}{t}=t^{\len{\lambda}} \delta_{\lambda\mu} \prod_{i\ge1}i^{m_i}m_i!,
  \end{align}
  where
  $m_i$ denotes of number of parts of $\lambda$ equal to $i$.
\end{enumerate}

For each $n\ge1$, the Jack polynomials in \(\fsym{n}\) may be defined as the images of the corresponding Jack functions in $\sym$ under the projection \(\pi_n\). As with monomial symmetric polynomials, we have
\(\pi_n\brac*{\jack{\lambda}{t}}=0\) if and only if \(\len{\lambda}>n\).  For \(\len{\lambda}\le n\), the Jack polynomials
\begin{align}
 \fjack{\lambda}{t}{z_1,\dots,z_n}=\pi_n(\jack{\lambda}{t})
\end{align}
are linearly independent and form a basis of \(\fsym{n}\).  For the application to follow, we mention the following important examples in $\fsym{n}$ called the \emph{rectangular} Jack polynomials.  In these, the partition has the form $\lambda = [m^n]$ in which all $n$ parts are equal to $m$.  Rectangular Jack polynomials have a very simple form:
\begin{equation} \label{eq:rectjack}
	\fjack{[m^n]}{t}{z_1,\dots,z_n} = \fmonsym{[m^n]}{z_1,\dots,z_n} = \prod_{i=1}^n z_i^m.
\end{equation}
This follows because all partitions of $mn$ that are strictly dominated by $[m^n]$ have length greater than $n$.  They also have extremely simple products with other Jacks.  For $\len{\lambda} \le n$, denote by $\lambda + [m^n]$ the partition with parts $\lambda_i + m$.  Then,
\begin{equation} \label{eq:rectjackpieri}
	\fjack{[m^n]}{t}{z_1,\dots,z_n} \fjack{\lambda}{t}{z_1,\dots,z_n} = \fjack{\lambda + [m^n]}{t}{z_1,\dots,z_n}.
\end{equation}
We emphasise that rectangular Jack polynomials are independent of the parameter $t$.

The Jack functions and polynomials satisfy many properties that shall be essential for what follows.  We list some of them here for convenience.
\begin{enumerate}
	\item We denote by $\djack{\lambda}{t}$ the elements of the basis dual to the $\jack{\lambda}{t}$ with respect to the inner product \eqref{innprod}. Since the Jack functions form an
  orthogonal basis, $\jack{\lambda}{t}$ is proportional to $\djack{\lambda}{t}$:
  \begin{align}
    \djack{\lambda}{t}=\nc{\lambda}{t}\jack{\lambda}{t}, \quad
    \nc{\lambda}{t} =\frac{1}{\cjprod{\jack{\lambda}{t},\jack{\lambda}{t}}{t}}.
  \end{align}
  The proportionality constant \(\nc{\lambda}{t}\) is given explicitly by
  \begin{align}\label{eq:djacknorm}
    \nc{\lambda}{t}=\prod_{s\in\lambda}\frac{a(s)t+l(s)+1}{(a(s)+1)t+l(s)},
  \end{align}
  where \(a(s)\) and \(l(s)\) denote the arm and leg lengths, respectively, of the box \(s\) in the Young diagram of \(\lambda\).
	\item The Jack functions and their duals admit a kind of generating function called the \emph{Cauchy kernel}:
  \begin{align}\label{cauchykernel}
    \prod_{i,j}(1-y_iz_j)^{-1/t}=
    \prod_{m\ge1}\exp\brac*{\frac{1}{t}\frac{\fpowsum{m}{y}\fpowsum{m}{z}}{m}}=
    \sum_{\lambda}\fjack{\lambda}{t}{y}\fdjack{\lambda}{t}{z}.
  \end{align}
  In this identity, the two alphabets $\set{y_i}$ and $\set{z_j}$ may be finite or infinite.
	\item Given partitions \(\lambda\) and \(\mu\), the skew Jack functions \(\jack{\lambda/\mu}{t}\) and \(\djack{\lambda/\mu}{t}\) are defined
	to be the unique symmetric functions satisfying
  \begin{equation}
    \cjprod{\jack{\lambda/\mu}{t},\djack{\nu}{t}}{t}=\cjprod{\jack{\lambda}{t},\djack{\mu}{t}\djack{\nu}{t}}{t}
    \qquad\text{and}\qquad
    \cjprod{\djack{\lambda/\mu}{t},\jack{\nu}{t}}{t}=\cjprod{\djack{\lambda}{t},\jack{\mu}{t}\jack{\nu}{t}}{t}
  \end{equation}
  for all partitions \(\nu\).  Let us write \(\mu\subseteq\lambda\) if the Young diagram of \(\mu\) is
  contained in that of \(\lambda\).  Then, \(\jack{\lambda/\mu}{t}= \djack{\lambda/\mu}{t}=0\) unless \(\mu\subseteq\lambda\).
  Finally, the ordinary and dual skew Jack functions are proportional:
  \begin{align}\label{skewdjack}
    \djack{\lambda/\mu}{t}=\frac{\nc{\lambda}{t}}{\nc{\mu}{t}}\jack{\lambda/\mu}{t}.
  \end{align}
	\item Consider an alphabet \(z=(z_1,z_2,\dots)\), partitioned into two subsets \(x=(x_1,x_2,\dots)\) and \(y=(y_1,y_2,\dots)\).  Any symmetric function in \(z\) may obviously be decomposed into symmetric functions in \(x\) and \(y\). For Jack functions, this decomposition is
  \begin{align} \label{unionjack}
    \fjack{\lambda}{t}{z}=\fjack{\lambda}{t}{x \cup y}=\sum_\nu \fjack{\nu}{t}{x}\fjack{\lambda/\nu}{t}{y}, \quad
    \fdjack{\lambda}{t}{z}=\fdjack{\lambda}{t}{x \cup y}=\sum_\nu \fdjack{\nu}{t}{x}\fdjack{\lambda/\nu}{t}{y}.
  \end{align}
  Both sums may clearly be restricted to partitions satisfying $\nu \subseteq \lambda$.
	\item The Jack polynomials $\fjack{\lambda}{t}{z_1,\dots,z_n}$ are orthogonal with respect to the inner product
	\begin{align} \label{eq:DefJProd}
		\jprod{f,g}{n}{t}=\int_{\cyc{n;t}} \fintker{n}{t}{x} f(x)\overline{g(x)}\,\frac{\dd x_1 \cdots \dd x_n}{x_1 \cdots x_n},
	\end{align}
	where \(\cyc{n;t}\) is the cycle normalised in \eqref{eq:cycnorm}, $\overline{g(x_1,x_2,\dots)}=g(x_1^{-1},x_2^{-1},\dots)$ and
	\begin{align}
		\fintker{n}{t}{x}=\prod_{1\le i\neq j\le n} \brac*{1-\frac{x_i}{x_j}}^{1/t}
	\end{align}
	is called the integrating kernel.  With respect to this integral inner product, the Jack polynomials satisfy
	\begin{align} \label{eq:DefNC} \jprod{\fjack{\lambda}{t}{x},\fdjack{\mu}{t}{x}}{n}{t}=\delta_{\lambda,\mu} \nc{\lambda}{t}(n), \quad
	  \nc{\lambda}{t}(n) = \prod_{s\in\lambda}\frac{n+a^\prime(s)t-l^\prime(s)}{n+(a^\prime(s)+1)t-l^\prime(s)-1},
	\end{align}
	where \(a^\prime(s)\) and \(l^\prime(s)\) denote the arm and leg colengths,
	respectively, of the box \(s\) in the Young diagram of \(\lambda\).
\end{enumerate}

\section{Explicit evaluation of $\Wvoa{3}$ singular vectors}
\label{sec:W3eval}

With this symmetric function technology under our belts, we now turn to the computation of singular vectors in Fock spaces, the idea being to realise them as images of \hwvs{} under a $\Wvoa{3}$-module homomorphism (screening operator).  For definiteness, we shall choose the screening operator $\scrs{+}{r_1,r_2}$ defined in \eqref{eq:scrs} that was constructed from $r_1$ copies of $\vo{\alpha_+ \alpha_1}$ and $r_2$ copies of $\vo{\alpha_+ \alpha_2}$.  The computation for $\scrs{-}{r_1,r_2}$ is exactly the same and will be omitted.  By \eqref{eq:welldefined}, $\scrs{+}{r_1,r_2}$ has a well defined action on the Fock space $\Fock{\eta}$, where $\eta = \svwts{r_1-r_2,s_1}{r_2,s_2}$, sending it into $\Fock{\theta}$, where $\theta = \svwts{-r_1,s_1}{r_1-r_2,s_2}$.

We can now explicitly evaluate the action of the screening operator $\scrs{+}{r_1,r_2}$ on the \hwv{} $\ket{\eta} \in \Fock{\eta}$.  Using \eqref{eq:scrfields}, \eqref{eq:mvfn} and \eqref{eq:defsvwts}, this action is
\begin{align}\label{integral}
	\scrs{+}{r_1,r_2} \ket{\eta}
	&= \int_{\Delta} \vop{\alpha_+\srt{1}}{z^1_1} \cdots \vop{\alpha_+\srt{1}}{z^1_{r_1}} \vop{\alpha_+ \srt{2}}{z^2_1} \cdots \vop{\alpha_+ \srt{2}}{z^2_{r_2}} \ket{\eta} \diff \nonumber \\
	&= \int_{\Delta} \dprod{k=1}{2}{1\le i\neq j\le r_k}{} \brac[\Big]{1 - \frac{z^k_i}{z^k_j}}^{\alpha_+^2}
		\cdot \dprod{i=1}{r_1}{j=1}{r_2} \brac[\Big]{1 - \frac{z^2_j}{z^1_i}}^{-\alpha_+^2}
		\cdot \dprod{k=1}{2}{i=1}{r_k} \brac[\big]{z^k_i}^{\alpha_+ \kil{\srt{k}}{\eta} + \alpha_+^2 (r_k-1) + 1} \nonumber\\
	&\qquad \cdot \prod_{i=1}^{r_1} \brac[\big]{z^1_i}^{-\alpha_+^2 r_2}
		\cdot \dprod{k=1}{2}{m\ge1}{} \exp \brac*{\frac{\alpha_+ a^k_{-m}}{m} \sum_{i=1}^{r_k} \brac[\big]{z^k_i}^m} \cdot \ket{\theta} \diffz \nonumber \\
	&= \int_{\Delta} \dprod{k=1}{2}{1\le i\neq j\le r_k}{} \brac[\Big]{1 - \frac{z^k_i}{z^k_j}}^{\alpha_+^2}
		\cdot \dprod{i=1}{r_1}{j=1}{r_2} \brac[\Big]{1 - \frac{z^2_j}{z^1_i}}^{-\alpha_+^2} \nonumber\\
	&\qquad \cdot \dprod{k=1}{2}{i=1}{r_k} \brac[\big]{z^k_i}^{s_k}
		\cdot \dprod{k=1}{2}{m\ge1}{} \exp \brac*{\frac{\alpha_+ a^k_{-m} \fpowsum{m}{z^k}}{m}} \cdot \ket{\theta} \diffz.
\end{align}
Here, the integrals are over the product cycle $\Delta = \cyc{r_1;\alpha_+^{-2}} \times \cyc{r_2;\alpha_+^{-2}}$, see \cref{sec:W3}.

To proceed, we note that the tensor product $\sym \otimes_{\CC} \sym$ is isomorphic to $\UEA{\heis_{-}} = \CC[a^k_{-m} \st k=1,2,\ m \in \ZZ_{>0}]$ as an algebra, by \eqref{eq:sym=polyring}. Concretely, let $y^1_i$ and $y^2_i$ denote the variables for the two factors of $\sym\otimes_{\CC}\sym$ and consider the isomorphism
\begin{align}\label{homom}
	\rho_+ \colon \Lambda \otimes_{\CC} \Lambda \lra \UEA{\heis_-}, \quad
	\powsum{m}(y^k) \longmapsto \frac{1}{\alpha_+} a^k_{-m}, \quad k=1,2,\ m \in \ZZ_{>0}.
\end{align}
Then, we may write
\begin{align}
	\prod_{m\ge1} \exp \brac*{\frac{\alpha_+ a^k_{-m} \fpowsum{m}{z^k}}{m}}
	= \rho_+ \brac*{\prod_{m\ge1} \exp \brac*{\alpha_+^2 \frac{\powsum{m}(y^k) \powsum{m}(z^k)}{m}}}
	= \rho_+ \brac*{\prod_{i\ge1} \prod_{j=1}^{r_k} (1-y^k_i z^k_j)^{-\alpha_+^2}},
\end{align}
recognising the Cauchy kernel \eqref{cauchykernel} with parameter $t = \alpha_+^{-2}$.

For $k=1$, we expand this Cauchy kernel in terms of Jacks and their duals as in \eqref{cauchykernel}:
\begin{equation} \label{eq:cauchyhom1}
	\prod_{m\ge1} \exp \brac*{\frac{\alpha_+ a^1_{-m} \fpowsum{m}{z^1}}{m}}
	= \rho_+ \brac*{\sum_\lambda \fjack{\lambda}{t}{y^1} \fdjack{\lambda}{t}{z^1}}.
\end{equation}
For $k=2$, we first combine the Cauchy kernel with that appearing in the second factor of the integrand of \eqref{integral}:
\begin{align} \label{eq:cauchyhom2}
	\rho_+ \brac*{\prod_{i\ge1} \prod_{j=1}^{r_2} (1-y^2_i z^2_j)^{-\alpha_+^2}}
	\prod_{i=1}^{r_1} \prod_{j=1}^{r_2} \brac*{1 - \brac{z^1_i}^{-1} z^2_j}^{-\alpha_+^2}
	= \rho_+ \brac*{\sum_\mu \fjack{\mu}{t}{y^2 \cup (z^1)^{-1}} \fdjack{\mu}{t}{z^2}}.
\end{align}
Here, we have noted that the product is a Cauchy kernel in the alphabets $\set{y^2_i} \cup \set{(z^1_\ell)^{-1}}$ and $\set{z^2_j}$.  This may be further simplified using skew-Jacks as in \eqref{unionjack}:
\begin{equation}
	\fjack{\mu}{t}{y^2 \cup (z^1)^{-1}} = \sum_{\nu} \overline{\fjack{\nu}{t}{z^1}} \fjack{\mu / \nu}{t}{y^2}.
\end{equation}
We recall that the skew-Jack $\jack{\mu / \nu}{t}$ is $0$ unless $\nu \subseteq \mu$.

Next, note that we also have the integrating kernels
\begin{equation}
	\dprod{k=1}{2}{1\le i\neq j\le r_k}{} \brac[\Big]{1 - \frac{z^k_i}{z^k_j}}^{\alpha_+^2} = \fintker{r_1}{t}{z^1} \fintker{r_2}{t}{z^2}
\end{equation}
of the symmetric polynomial inner product \eqref{eq:DefJProd}.  Finally, the product $\dprod{k=1}{2}{i=1}{r_k} \brac[\big]{z^k_i}^{s_k}$ is a product of rectangular Jack polynomials.  However, here we have to be careful with the signs of the $s_k$.  Indeed, \eqref{eq:DefConfWt} and \eqref{eq:defsvwts} show that the conformal weights of the \hwvs{} $\ket{\eta}$ and $\ket{\theta}$ differ by
\begin{equation} \label{eq:DeltaH}
	h_{\eta} - h_{\theta} = -r_1 s_1 - r_2 s_2.
\end{equation}
This must be non-negative if the screening operator $\scrs{+}{r_1,r_2}$ is to map $\ket{\eta}$ to a singular descendant of $\ket{\theta}$.  We shall therefore assume from here on that $s_1, s_2 \in \ZZ_{\le 0}$.  Thus,
\begin{equation}
	\dprod{k=1}{2}{i=1}{r_k} \brac[\big]{z^k_i}^{s_k} = \overline{\fjack{[-s_1^{r_1}]}{t}{z^1}} \, \overline{\fjack{[-s_2^{r_2}]}{t}{z^2}},
\end{equation}
using \eqref{eq:rectjack}.

Putting all this back in \eqref{integral}, the integrand factorises and we get
\begin{align}\label{eq:sveval}
\scrs{+}{r_1,r_2} \ket{\eta}
	&= \sum_{\lambda, \mu, \nu} \rho_+ \brac*{\fjack{\lambda}{t}{y^1}}
		\int_{\cyc{r_1;t}} \fintker{r_1}{t}{z^1} \overline{\fjack{[-s_1^{r_1}]}{t}{z^1}} \, \overline{\fjack{\nu}{t}{z^1}}
		\fdjack{\lambda}{t}{z^1} \prod_{i=1}^{r_1} \frac{\dd z^1_i}{z^1_i} \notag \\
	&\mspace{80mu} \cdot \rho_+ \brac*{\fjack{\mu / \nu}{t}{y^2}}
		\int_{\cyc{r_2;t}} \fintker{r_2}{t}{z^2} \overline{\fjack{[-s_2^{r_2}]}{t}{z^2}} \fdjack{\mu}{t}{z^2} \prod_{i=1}^{r_2} \frac{\dd z^2_i}{z^2_i} \cdot \ket{\theta} \notag \\
	&= \sum_{\lambda, \mu, \nu} \jprod{\djack{\lambda}{t}, \jack{\nu + [-s_1^{r_1}]}{t}}{r_1}{t}
		\jprod{\djack{\mu}{t}, \jack{[-s_2^{r_2}]}{t}}{r_2}{t}
		\rho_+ \brac*{\fjack{\lambda}{t}{y^1} \fjack{\mu / \nu}{t}{y^2}} \ket{\theta} \notag \\
	&= \sum_{\substack{\nu \subseteq [-s_2^{r_2}]\\\ell(\nu)\le r_1}} \nc{\nu + [-s_1^{r_1}]}{t}(r_1) \nc{[-s_2^{r_2}]}{t}(r_2)
		\rho_+ \brac*{\fjack{\nu + [-s_1^{r_1}]}{t}{y^1} \fjack{[-s_2^{r_2}] / \nu}{t}{y^2}} \ket{\theta},
\end{align}
by using \eqref{eq:rectjackpieri}, \eqref{eq:DefJProd} and \eqref{eq:DefNC}.  As the second factor is independent of $\nu$ (and non-zero), it may be absorbed into the normalisation of the \sv{}.  Our final result is therefore
\begin{equation} \label{eq:W3SVFormula}
	\scrs{+}{r_1,r_2} \ket{\eta} = \sum_{\substack{\nu \subseteq [-s_2^{r_2}]\\\ell(\nu)\le r_1}} \nc{\nu + [-s_1^{r_1}]}{t}(r_1) \rho_+ \brac*{\fjack{\nu + [-s_1^{r_1}]}{t}{y^1} \fjack{[-s_2^{r_2}] / \nu}{t}{y^2}} \ket{\theta}.
\end{equation}
This form is now easily implemented in computer algebra packages.

The \rhs{} of \eqref{eq:W3SVFormula} is easily seen to be manifestly non-zero
by noting that the total degree, with respect to the \(a^2_{-m}\), of the summand corresponding to the empty partition
\(\nu=\sqbrac*{\ }\) is maximal and that all other summands have strictly lesser degrees.  Since \(\nc{[-s_1^{r_1}]}{t}(r_1)\), \(\fjack{[-s_1^{r_1}]}{t}{y^1}\) and \(\fjack{[-s_2^{r_2}]}{t}{y^2}\)
are all non-zero, this summand is therefore linearly independent of all others. The conclusion is that \eqref{eq:W3SVFormula} defines a \sv{} for every $r_1, r_2 \in \ZZ_{\ge 0}$, $s_1, s_2 \in \ZZ_{\le0}$ and $t \in \CC \setminus \QQ_{\le 0}$.

\section{Examples}
\label{sec:examples}

We now illustrate the $\wvoa{}$ \sv{} formula \eqref{eq:W3SVFormula} with three examples.

\subsection{Example 1} \label{ex:1}

For our first example, we compute a singular vector for the case when $t=\frac{4}{5}$, so that
\begin{equation}
	\alpha_+ = \frac{\sqrt{5}}{2}, \quad \alpha_- = -\frac{2}{\sqrt{5}}, \quad \alpha_0 = \frac{1}{2 \sqrt{5}}, \quad c = \frac{4}{5}.
\end{equation}
This central charge corresponds to that of the $3$-state Potts model, described by the $\wvoa$ minimal model $\wminmod{4,5}$ (the parameters here are the numerator and denominator of $t$ in reduced form).  Take $r_1=r_2=-s_1=-s_2=1$ for simplicity.  Then, the map $\scrs{+}{1,1}$ sends $\Fock{\eta}$, where $\eta=\svwts{0,-1}{1,-1}$, into $\Fock{\theta}$, where $\theta=\svwts{-1,-1}{0,-1}$.  We note that
\begin{equation} \label{eq:ex1hw}
	h_{\eta} = \frac{13}{6}, \quad h_{\theta} = \frac{1}{6}, \quad
	w_{\eta} = \frac{187}{9 \sqrt{390}}, \quad w_{\theta} = -\frac{7}{9 \sqrt{390}},
\end{equation}
by \eqref{eq:DefConfWt} and \eqref{eq:DefWWt}.  The conformal weight $h_{\theta}$ is not one of those associated with the $3$-state Potts model.  Nevertheless, the Fock space $\Fock{\theta}$ has a singular vector at grade 2 in accordance with \eqref{eq:DeltaH}.  \cref{eq:W3SVFormula} writes it in the form
\begin{align}
	\scrs{+}{1,1} \ket{\eta} &= \sum_{\nu \subseteq [1]} \nc{\nu + [1]}{4/5}(1) \rho_+ \brac*{\fjack{\nu + [1]}{4/5}{y^1} \fjack{[1] / \nu}{4/5}{y^2}} \ket{\theta}.
\end{align}

There are only two partitions $\nu$ to consider.  Using \eqref{eq:DefNC}, \eqref{homom} and \textsc{SageMath} to write Jacks and skew-Jacks in terms of power sums, we have
\begin{equation}
  \begin{aligned}
    \nu &= [0]: & \nc{[0]+[1]}{4/5}(1) &= \frac{5}{4}, &
    \rho_+ \brac*{\fjack{[0]+[1]}{4/5}{y^1}} &= \rho_+ \brac*{\fpowsum{[1]}{y^1}}=\frac{2}{\sqrt{5}}a^{1}_{-1}, \\
    &&&& \rho_+ \brac*{\fjack{[1]/[0]}{4/5}{y^2}} &= \rho_+ \brac*{\fpowsum{[1]}{y^2}}=\frac{2}{\sqrt{5}}a^{2}_{-1}. \\
    \nu &= [1]: & \nc{[1]+[1]}{4/5}(1) &= \frac{5}{4} \frac{9}{8}, &
    \rho_+ \brac*{\fjack{[1]+[1]}{4/5}{y^1}} &= \rho_+ \brac*{\frac{5}{9}\fpowsum{[1,1]}{y^1}+\frac{4}{9}\fpowsum{[2]}{y^1}} \\
    &&&&&= \frac{4}{5} \frac{5}{9} a^1_{-1} a^1_{-1} + \frac{2}{\sqrt{5}} \frac{4}{9} a^1_{-2}, \\
    &&&& \rho_+ \brac*{\fjack{[1]/[1]}{4/5}{y^1}} &= \rho_+ \brac*{\fpowsum{[0]}{y^1}}=1.
  \end{aligned}
\end{equation}
The singular vector is therefore explicitly identified as
\begin{align} \label{eq:ex1sv}
	\scrs{+}{1,1} \ket{\eta}=\brac*{a^1_{-1}a^2_{-1}+\frac{5}{8}a^1_{-1}a^1_{-1}+\frac{\sqrt{5}}{4}a^1_{-2}}\ket{\theta}.
\end{align}

Consider the $\wvoa$ Verma module $\Verma{\vartheta}$ whose \hwv{} $\ket{\vartheta}$ has $L_0$- and $W_0$-eigenvalue $h_{\theta}$ and $w_{\theta}$, as given in \eqref{eq:ex1hw}. By direct calculation, $\Verma{\vartheta}$ has a singular vector $\ket{\chi}$, unique up to normalisation, at grade 2:
\begin{align}
	\ket{\chi} = \brac*{\frac{390}{119} W_{-1}W_{-1} - \frac{\sqrt{390}}{17} W_{-2} + \frac{10\sqrt{390}}{119} L_{-1}W_{-1} + L_{-1}L_{-1}} \ket{\vartheta}.
\end{align}
The free field realisation $f \colon \wvoa \ira \hvoa{2}$ defined by \eqref{eq:ffr} induces a $\wvoa{}$-module homomorphism
\begin{equation}
	f_{\vartheta} \colon \Verma{\vartheta} \lra \Fock{\theta}, \quad f_{\vartheta}(U \ket{\vartheta}) = f(U) \ket{\theta}.
\end{equation}
Here, $U$ is an arbitrary element of the $\wvoa$ mode algebra, this being the (unital) associative algebra generated by the $L_m$ and $W_n$ subject to \eqref{eq:W3ModeAlgebra}.  Explicit calculation now verifies that the image of the singular vector $\ket{\chi}$ under $f_{\vartheta}$ is, of course, that constructed in \eqref{eq:ex1sv}:
\begin{align}
f_{\vartheta} \brac*{\ket{\chi}} = \brac*{\frac{5}{4} a^1_{-1} a^2_{-1} + \frac{25}{32} a^1_{-1} a^1_{-1} + \frac{5 \sqrt{5}}{16} a^1_{-2}} \ket{\theta} =\frac{5}{4} \scrs{+}{1,1} \ket{\eta}.
\end{align}

\subsection{Example 2}

For our second example, we compute a grade three singular vector for general central charges.
Let $r_1=2$ and $r_2=-s_1=-s_2=1$, so that $\eta = \svwts{1,-1}{1,-1}$ and $\theta = \svwts{-2,-1}{1,-1}$.
In order to evaluate the singular vector formula \eqref{eq:W3SVFormula}, we need to compute
\begin{equation}
	\begin{gathered}
		\nc{[1,1]}{t}(2) = \frac{2}{t+1} \frac{1}{t}, \quad
		\nc{[2,1]}{t}(2) = \frac{2}{t+1} \frac{1}{t} \frac{t+2}{2t+1}, \\
		\jack{[1,1]}{t} = \frac{1}{2} \powsum{[1,1]} - \frac{1}{2} \powsum{[2]}, \quad
		\jack{[1]/[0]}{t} = \powsum{[1]}, \\
		\jack{[2,1]}{t} = \frac{1}{t+2} \powsum{[1,1,1]} + \frac{t-1}{t+2} \powsum{[2,1]} - \frac{t}{t+2} \powsum{[3]}, \quad
		\jack{[1]/[1]}{t} = 1,
	\end{gathered}
\end{equation}
again using \eqref{eq:DefNC}, \eqref{homom} and \textsc{SageMath}. The singular vector is thus
\begin{align}
	\scrs{+}{2,1} \ket{\eta} &= \sqbrac*{
		\nc{[1,1]}{t}(2) \rho_+ \brac*{\fjack{[1,1]}{t}{y^1} \fjack{[1]/[0]}{t}{y^2}} +
		\nc{[2,1]}{t}(2) \rho_+ \brac*{\fjack{[2,1]}{t}{y^1} \fjack{[1]/[1]}{t}{y^2}}
              } \ket{\theta} \notag \\
	&= \left[
		\frac{1}{t(t+1)} \brac*{\frac{1}{\alpha_+^2} a^1_{-1} a^1_{-1} - \frac{1}{\alpha_+} a^1_{-2}} \frac{1}{\alpha_+} a^2_{-1} \right. \notag \\
		&\qquad \left. + \frac{2}{t(t+1)(2t+1)} \brac*{\frac{1}{\alpha_+^3} a^1_{-1} a^1_{-1} a^1_{-1} + \frac{t-1}{\alpha_+^2} a^1_{-2} a^1_{-1} - \frac{t}{\alpha_+} a^1_{-3}}
	\right] \ket{\theta} \notag \\
	&= \left[
		\frac{2/\alpha_+}{(t+1)(2t+1)} a^1_{-1} a^1_{-1} a^1_{-1} + \frac{1/\alpha_+}{t+1} a^1_{-1} a^1_{-1} a^2_{-1}	\right. \notag \\
		&\qquad \left. + \frac{2(t-1)}{(t+1)(2t+1)} a^1_{-2} a^1_{-1} - \frac{1}{t+1} a^1_{-2} a^2_{-1} - \frac{1/\alpha_+}{(t+1)(2t+1)} a^1_{-3}
	\right] \ket{\theta}.
\end{align}
We note that the result is manifestly well defined and non-zero for all $\alpha_+$ such that $t = \alpha_+^{-2} \in \CC \setminus \QQ_{\le 0}$, as expected.  This region includes all central charges less than $98$.

\subsection{Example 3}

Our final example concerns \svs{} for quite arbitrary central charges (including all $c < 98$).  This time, we fix $\theta = 0$ and use \eqref{eq:W3SVFormula} to construct \svs{} in the Fock space $\Fock{0}$.

First, we note that $\theta = \svwts{-r_1,s_1}{r_1-r_2,s_2} = 0$ may be solved for $r_1$ and $s_1$:
\begin{equation}
	r_1 = -1 + (-s_1 + 1) t, \quad r_2 = -2 + (-s_1 - s_2 + 2) t.
\end{equation}
Since $r_1, r_2, s_1, s_2 \in \ZZ$, we will only find \svs{} when $t \in \QQ_{>0}$.  Writing $t = \frac{u}{v}$, where $u$ and $v$ are coprime integers, it follows that
\begin{equation}
	r_1 = mu-1, \quad -s_1 = mv-1, \quad r_2 = nu-2, \quad -s_2 = (n-m)v-1,
\end{equation}
for some $m,n \in \ZZ$.  Given that $r_1, r_2 \in \ZZ_{\ge 0}$ and $s_1, s_2 \in \ZZ_{<0}$, we conclude that $m$, $n$ and $n-m$ must be positive integers.  We thereby obtain, for each fixed $t \in \QQ_{>0}$, an infinite sequence of \svs{}, generically indexed by integers $n>m>0$, of the form $\scrs{+}{mu-1,nu-2} \ket{\svwts{(m-n)u+1,-mv+1}{nu-2,-(n-m)v+1}}$.  Among these, the \sv{} of lowest grade corresponds, assuming that $u>1$, to $(m,n) = (1,2)$.  Moreover, the grade of $\scrs{+}{u-1,2(u-1)} \ket{\svwts{-(u-1),-(v-1)}{2(u-1),-(v-1)}}$ is $3(u-1)(v-1)$, by \eqref{eq:DeltaH}.

It is not clear if these \svs{} of the Fock space $\Fock{0}$ correspond, in the sense of Example~\ref{ex:1} to \svs{} in the $\wvoa$ vacuum Verma module $\Verma{0}$ or not.  However, there are five other Fock spaces $\Fock{\zeta}$ whose \hwvs{} $\ket{\zeta}$ have $h_{\zeta} = w_{\zeta} = 0$.  This follows from the easily verified fact that both $h_{\zeta}$ and $w_{\zeta}$ are left invariant by the following shifted action of the Weyl group $\symgp{3}$:
\begin{equation}
	\sigma \cdot \zeta = \sigma(\zeta - \alpha_0 \weyl) + \alpha_0 \weyl, \quad \sigma \in \symgp{3}.
\end{equation}
Each of these five other Fock spaces has an infinite sequence of \svs{} given by \eqref{eq:W3SVFormula} and it is interesting to ask whether these also correspond to \svs{} in $\Verma{0}$ or not.  We shall not investigate this question here.  We only note the following observation:  $\Fock{2 \alpha_0 \weyl}$ has such a \sv{} at grade $3$ and it corresponds to just one of the \emph{two} linearly independent grade $3$ \svs{} of $\Verma{0}$.  Which one is obtained depends on the branch of the square root of $\beta$ chosen in \eqref{eq:ffr}.

We conclude by remarking that the question of whether the Fock space \svs{} constructed here exhaust the \svs{} of $\Verma{0}$ is much easier to answer.  They do not.  We cannot obtain the two linearly independent \svs{} at grade $1$ using \eqref{eq:W3SVFormula} (for $c \neq 2$; when $c=2$, $W_0$ acts non-diagonalisably).  Nor can we obtain, when $t = \frac{u}{v} \in \QQ_{>0}$, the grade $(u-2)(v-2)$ \sv{} whose image is non-zero in the universal $\wvoa$ vacuum module.  This \sv{} can be constructed formally using screening operators, but we do not know how to actually evaluate the integral in this case.  What is needed is a certain $\SLA{sl}{3}$ analogue of the theory of Jack functions, something which does not appear to have yet been developed (see \cite{TarSel03,WarSel09,WarSL310} for work in this direction).  We hope to return to this important point in the future.

\section{Explicit evaluation of $\Wvoa{N}$ singular vectors}
\label{sec:WN}

In the previous section, we computed explicit formulae for
$\Wvoa{3}$ singular vectors in Fock spaces. In this section we generalise the results of
\cref{sec:W3eval} and derive explicit formulae for $\Wvoa{N}$ singular vectors in Fock spaces.
Continuing the pattern of ranks $1$ and $2$, the rank $N-1$ Heisenberg vertex
operator algebra, with choice of energy-momentum tensor \eqref{eq:emtensor},
has $2(N-1)$ screening operators, $\vop{\zeta}{w}$ for $\zeta =
\alpha_\pm \srt{i}$, where $\alpha_\pm$ was defined in \eqref{eq:alphadef} and the $\srt{i}$ are the simple roots of $\alg{sl}(N)$.

The \Walg{N} vertex operator algebra $\Wvoa{N}$ is usually described as being generated by the Virasoro field
$T(z)$ and $N-2$ Virasoro primary fields $W^3(z),\dots,W^{N}(z)$
of conformal weights $3,\dots,N$, respectively. Unfortunately, explicit formulae for these primaries, for example in terms of Heisenberg fields, rapidly increase in complexity as \(N\) increases
and there are no known closed formulae for general \(N\). Fortunately our computations do not require explicit expressions for the $W^k(z)$, only the fact that they commute with the screening operators.

We therefore turn to the definition of the \Walg{N} vertex operator algebra \cite{Luk88} in terms of a generating function called the quantum Miura transform.
This constructs a different set of generators of $\Wvoa{N}$ that are not conformal primaries in general, but which are easily verified to commute with screening operators.
We denote these new generating fields by $U^2(z)=T(z)$, $U^3(z),\dots,U^{N}(z)$ and their generating function by
\begin{align} \label{eq:Miura}
  R_{N}(z)&=-\sum^{N}_{k=0}U_k(z)(\alpha_0\partial)^{N-k}
  =\normord{(\alpha_0\partial_z-\epsilon^1(z))\cdots(\alpha_0\partial_z-\epsilon^{N}(z))},
\end{align}
where the \(\epsilon^i\) are the weights of the defining representation of \(\alg{sl}(N)\) so that $\epsilon^1 + \cdots + \epsilon^N = 0$ and \(\alpha^i=\epsilon^{i}-\epsilon^{i+1}\), for \(i=1,\dots,N-1\).

With the \Walg{N} algebra now defined explicitly as the algebra generated by the \(U^i(z)\), \(i=2,\dots N\), we construct screening fields in a manner similar to $\Wvoa{3}$. As mentioned above, the
vertex operators $\vop{\zeta}{w}$ with Heisenberg weights $\zeta=\alpha_{\pm}\alpha_1,\dots,\alpha_{\pm}\alpha_{N-1}$ are screening fields, because their operator product expansions with $R_N(z)$ are total derivatives:
\begin{equation} \label{eq:miuraope}
	R_{N}(z)\vop{\alpha_\pm\alpha_i}{w}\sim \partial_w\brac*{\frac{\normord{R^i_{N}(w)\vop{\alpha_\pm\alpha_i}{w}}}{z-w}}.
\end{equation}
Here, \(R^i_{N}(z)\) is defined as the product in \eqref{eq:Miura}, but without the factors involving \(\epsilon^i\) and \(\epsilon^{i+1}\).

As in the rank $2$ case, the residues of the screening fields, when defined, commute with the \Walg{N} algebra, because their operator product
expansions with the generating \(U^i\) fields are total derivatives, and therefore
define module homomorphisms. Also as in the rank 2 case, one can compose
screening fields and integrate them over suitable contours to construct yet
more module homomorphisms.
Note that it is sufficient to only compose screening fields whose weights are all rescalings of simple \(\alg{sl}(N)\) roots by either
\(\alpha_+\) or \(\alpha_-\). This is because the two screening operators corresponding to the residues of $\vop{\alpha_+\alpha^i}{w}$ and $\vop{\alpha_-\alpha^j}{w}$
commute and can thus be considered independently. We shall therefore only present
calculations involving the $\vop{\alpha_+\alpha^i}{w}$; those involving the $\vop{\alpha_-\alpha^j}{w}$ work in exactly the same way.

We therefore compose $r_1$ copies of $\vop{\alpha_+\alpha^1}{z^1}$ with $r_{2}$ copies of $\vop{\alpha_+\alpha^{2}}{z^{2}}$ and so on, evaluating this composition on a Fock space of weight $\eta$, to obtain
\begin{align}
	&\Bigl.\prod_{i=1}^{r_1} \vop{\alpha_+ \srt{1}}{z^1_i} \cdots \prod_{i=1}^{r_{N-1}} \vop{\alpha_+ \srt{N-1}}{z^{N-1}_i} \Bigr\rvert_{\Fock{\eta}}\nonumber\\
	&\quad = \prod^{N-1}_{k=1} \prod_{1\le i<j\le r_k} (z^k_i-z^k_j)^{2\alpha_+^2} \cdot \prod^{N-2}_{k=1} \prod_{i=1}^{r_k} \prod_{j=1}^{r_{k+1}} (z^k_i-z^{k+1}_j)^{-\alpha_+^2} \cdot \prod_{k=1}^{N-1} \prod_{i=1}^{r_k} (z^k_i)^{\alpha_+\kil{\srt{k}}{\eta}}\nonumber\\
	&\quad\quad \cdot \prod_{k=1}^{N-1} \gae{r_k \alpha_+ \srt{k}}
	\cdot \prod_{k=1}^{N-1} \prod_{m\ge1} \exp \brac*{\frac{\alpha_+ \scrt{k}_{-m}}{m} \sum_{i=1}^{r_k} (z_i^k)^m} \exp \brac*{-\frac{\alpha_+ \scrt{k}_{m}}{m} \sum_{i=1}^{r_k} (z_i^k)^{-m}} \nonumber\\
	&\quad = \prod^{N-1}_{k=1} \prod_{1\le i\neq j\le r_k} \brac[\Big]{1-\frac{z_i^k}{z^k_j}}^{\alpha_+^2} \cdot \prod^{N-1}_{k=2} \prod_{i=1}^{r_k} \prod_{j=1}^{r_{k+1}} \brac[\Big]{1-\frac{z^k_j}{z^{k-1}_i}}^{-\alpha_+^2} \cdot \prod_{k=1}^{N-1} \prod_{i=1}^{r_k} (z^k_i)^{\alpha_+^2 (r_k-r_{k+1}-1) + \alpha_+ \kil{\srt{k}}{\eta}}\nonumber\\
	&\quad\quad\cdot \prod_{k=1}^{N-1} \gae{r_k \alpha_+ \srt{k}}
	\cdot \prod_{k=1}^{N-1} \prod_{m\ge1} \exp \brac*{\frac{\alpha_+ \scrt{k}_{-m}}{m} \sum_{i=1}^{r_k} (z_i^k)^m} \exp \brac*{-\frac{\alpha_+ \scrt{k}_{m}}{m} \sum_{i=1}^{r_k} (z_i^k)^{-m}},
\end{align}
where we define \(r_{N}=0\).
In analogy to the reasoning presented for the \(\Wvoa{3}\) algebra in \cref{sec:W3}, one can construct a $\Wvoa{N}$-module homomorphism by choosing an appropriate contour. Integrating over the contours of Tsuchiya and Kanie \cite{TKScr86} is well defined whenever
\begin{equation}
 \label{eq:allowedweights}
 \alpha_+^2\brac*{r_{k}-r_{k+1}-1}+\alpha_+\kil{\srt{k}}{\eta}\in \ZZ,\
 \text{for all}\ k=1,\dots,N-1.
\end{equation}
To parametrise the weights satisfying these constraints, we define
\begin{align}
  \zeta_{\vec{u}, \vec{v}}
  =\sum^{N-1}_{i=1}\brac*{(1-u_i)\alpha_++(1-v_i)\alpha_-}\fwt{i},  \quad
  \vec{u}=(u_1,\dots,u_{N-1}),\ \vec{v}=(v_1,\dots,v_{N-1})\in \ZZ^{N-1},
\end{align}
and define screening operators
\begin{align}
\scrs{+}{\vec{r}} &= \int_{\cyc{r_1;1/\alpha_+^2}} \cdots \int_{\cyc{r_{N-1};1/\alpha_+^2}}
\prod_{i=1}^{r_1}\vop{\alpha_\pm\srt{1}}{z^1_i} \cdots \prod_{i=1}^{r_{N-1}} \vop{\alpha_+ \srt{N-1}}{z^{N-1}_i} \cdot \diffn,
\quad \vec{r} \in \ZZ_{\ge 0}^{N-1}.
\end{align}
These, in turn, induce $\Wvoa{N}$-module homomorphisms
\begin{align}
	\scrs{+}{\vec{r}} \colon \Fock{\eta^+_{\vec{r},\vec{s}}} &\to \Fock{\theta^+_{\vec{r},\vec{s}}}, \quad \vec{r} \in \ZZ_{\ge 0}^{N-1},\ \vec{s} \in \ZZ^{N-1},
\end{align}
where $\eta^+_{\vec{r},\vec{s}} = \zeta_{(r_1-r_2,\dots,r_{N-2}-r_{N-1},r_{N-1}), \vec{s}}$ and $\theta^+_{\vec{r},\vec{s}} = \zeta_{(-r_1,r_1-r_2,\dots,r_{N-2}-r_{N-1}), \vec{s}}$.  Similar screening operators $\scrs{-}{\vec{s}}$ are obtained by swapping the roles of $\alpha_+$ and $\alpha_-$, as well as $\vec{r}$ and $\vec{s}$, in this development.

If we apply the screening operator $\scrs{+}{\vec{r}}$ to the $\hvoa{N-1}$ \hwv{} $\ket{\eta^+_{\vec{r},\vec{s}}}$, we get
\begin{align}\label{integralWN}
	\scrs{+}{\vec{r}} \ket{\eta^+_{\vec{r},\vec{s}}}
	&= \int_{\cyc{r_1;1/\alpha_+^2}} \cdots \int_{\cyc{r_{N-1};1/\alpha_+^2}} \prod_{i=1}^{r_1} \vop{\alpha_+ \srt{1}}{z^1_i} \cdots \prod_{i=1}^{r_{N-1}} \vop{\alpha_+ \srt{N-1}}{z^{N-1}_i} \cdot \ket{\eta^+_{\vec{r},\vec{s}}} \diffn \nonumber \\
	&= \int_{\cyc{r_1;1/\alpha_+^2}} \cdots \int_{\cyc{r_{N-1};1/\alpha_+^2}} \dprod{k=1}{N-1}{1\le i\neq j\le r_k}{} \brac[\Big]{1 - \frac{z^k_i}{z^k_j}}^{\alpha_+^2}
	\cdot \prod_{k=2}^{N-1} \dprod{i=1}{r_{k-1}}{j=1}{r_{k}} \brac[\Big]{1 - \frac{z^k_j}{z^{k-1}_i}}^{-\alpha_+^2}
	 \nonumber\\
	&\quad \cdot \dprod{k=1}{N-1}{i=1}{r_k} \brac[\big]{z^k_i}^{\alpha_+ \kil{\srt{k}}{\eta^+_{\vec{r},\vec{s}}} + \alpha_+^2 (r_k-1) + 1}\cdot \dprod{k=2}{N-1}{i=1}{r_k} \brac[\big]{z^1_i}^{-\alpha_+^2 r_k}\nonumber\\
	&\quad\quad\cdot \dprod{k=1}{N-1}{m\ge1}{} \exp \brac*{\frac{\alpha_+ a^k_{-m}}{m} \sum_{i=1}^{r_k} \brac[\big]{z^k_i}^m} \cdot \ket{\theta^+_{\vec{r},\vec{s}}} \diffnz \nonumber \\
	&= \int_{\cyc{r_1;1/\alpha_+^2}} \cdots \int_{\cyc{r_{N-1};1/\alpha_+^2}} \dprod{k=1}{N-1}{1\le i\neq j\le r_k}{} \brac[\Big]{1 - \frac{z^k_i}{z^k_j}}^{\alpha_+^2}
	\cdot \prod_{k=1}^{N-2}\dprod{i=1}{r_{k-1}}{j=1}{r_{k}} \brac[\Big]{1 - \frac{z^k_j}{z^{k-1}_i}}^{-\alpha_+^2} \nonumber\\
	&\qquad \cdot \dprod{k=1}{N-1}{i=1}{r_k} \brac[\big]{z^k_i}^{s_k}
	\cdot \dprod{k=1}{N-1}{m\ge1}{} \exp \brac*{\frac{\alpha_+ a^k_{-m} \fpowsum{m}{z^k}}{m}} \cdot \ket{\theta^+_{\vec{r},\vec{s}}} \diffnz.
\end{align}
As in the  \(\Wvoa{3}\) case, we can evaluate these integral formulae for
singular vectors in terms of symmetric functions. Recall that the tensor product of $N-1$ copies of the ring of symmetric functions
$\sym^{\otimes N-1}$ is isomorphic to
\begin{equation}
	\UEA{\heis_{-}} = \CC[a^k_{-m} \st k=1,\dots,N-1,\ m \in \ZZ_{>0}]
\end{equation}
as an algebra, by
\eqref{eq:sym=polyring}. Distinguishing the alphabets of the tensor factors by superscripts, so that the alphabet in the \(i\)-th tensor factor is denoted by \(y^i\), we define the following algebra isomorphism generalising that of \eqref{homom}:
\begin{align}
\rho_+ \colon \sym^{\otimes N-1} \rightarrow \UEA{\heis_-^{\otimes N-1}}, \quad
\fpowsum{n}{y^k} \mapsto \frac{1}{\alpha_+} a^k_{-n}.
\end{align}
This isomorphism allows us to write
\begin{align}
\prod_{m\ge1} \exp \brac*{\frac{\alpha_+ a^k_{-m} \fpowsum{m}{z^k}}{m}}
= \rho_+ \brac*{\prod_{m\ge1} \exp \brac*{\alpha_+^2 \frac{\powsum{m}(y^k) \powsum{m}(z^k)}{m}}}
= \rho_+ \brac*{\prod_{i\ge1} \prod_{j=1}^{r_k} (1-y^k_i z^k_j)^{-\alpha_+^2}},
\end{align}
for $k=1,\dots,N-1$.  We now identify, with $t = \alpha_+^{-2}$,
\begin{align}
\dprod{k=1}{N-1}{1\le i\neq j\le r_k}{} \brac[\Big]{1 - \frac{z^k_i}{z^k_j}}^{\alpha_+^2}=\prod_{k=1}^{N-1}\fintker{r_k}{t}{z^k}
\end{align}
as the product of the integrating kernels for the variables $z_k$. For $k=1$, as in \eqref{eq:cauchyhom1}, we write
\begin{equation}
\prod_{m\ge1} \exp \brac*{\frac{\alpha_+ a^1_{-m} \fpowsum{m}{z^1}}{m}}
= \rho_+ \brac*{\sum_{\mu_1} \fjack{\mu_1}{t}{y^1} \fdjack{\mu_1}{t}{z^1}}.
\end{equation}
For $k=2,\dots,N-1$, similar to \eqref{eq:cauchyhom2}, we have instead
\begin{align}
\prod_{m\ge1} \exp \brac*{\frac{\alpha_+ a^k_{-m} \fpowsum{m}{z^k}}{m}} \cdot \dprod{i=1}{r_{k-1}}{j=1}{r_{k}} \brac[\Big]{1 - \frac{z^{k}_j}{z^{k-1}_i}}^{-\alpha_+^2}
&=\rho_+ \brac*{\sum_{\mu_k}  \fjack{\mu_k}{t}{y^k \cup (z^{k-1})^{-1}}\fdjack{\mu_k}{t}{z^k}}\nonumber\\
&= \rho_+ \brac*{\sum_{\mu_k,\nu_k}\overline{\fjack{\nu_k}{t}{z^{k-1}}}\fjack{\mu_k / \nu_k}{t}{y^k} \fdjack{\mu_k}{t}{z^k}}.
\end{align}

Putting everything together, we have
\begin{align}
\scrs{+}{\vec{r}} \ket{\eta^+_{\vec{r},\vec{s}}}
&= \sum_{\substack{\mu_1,\mu_2,\dots, \mu_{N-1} \\ \hphantom{\mu_1,} \nu_2,\dots,\nu_{N-1}}} \rho_+ \brac*{\fjack{\mu_1}{t}{y^1}}
\int_{\cyc{r_1;t}} \fintker{r_1}{t}{z^1} \overline{\fjack{[-s_1^{r_1}]}{t}{z^1}} \, \overline{\fjack{\nu_2}{t}{z^1}}
\fdjack{\mu_1}{t}{z^1} \prod_{i=1}^{r_1} \frac{\dd z^1_i}{z^1_i} \notag \\
&\quad\cdot \rho_+ \brac*{\fjack{\mu_2 / \nu_2}{t}{y^2}} \int_{\cyc{r_2;t}} \fintker{r_2}{t}{z^2} \overline{\fjack{[-s_2^{r_2}]}{t}{z^2}} \, \overline{\fjack{\nu_3}{t}{z^2}}
\fdjack{\mu_2}{t}{z^2} \prod_{i=1}^{r_2} \frac{\dd z^2_i}{z^2_i} \notag \\
&\mspace{250mu}\vdots\notag\\
&\quad \cdot \rho_+ \brac*{\fjack{\mu_{N-2} / \nu_{N-2}}{t}{y^{N-2}}}
\int_{\cyc{r_{N-2};t}} \fintker{r_{N-2}}{t}{z^{N-2}} \overline{\fjack{[-s_{N-2}^{r_{N-2}}]}{t}{z^{N-2}}} \, \overline{\fjack{\nu_{N-1}}{t}{z^{N-2}}} \fdjack{\mu_{N-2}}{t}{z^{N-2}} \prod_{i=1}^{r_{N-2}} \frac{\dd z^{N-2}_i}{z^{N-2}_i} \notag \\
&\quad \cdot \rho_+ \brac*{\fjack{\mu_{N-1} / \nu_{N-1}}{t}{y^{N-1}}}
\int_{\cyc{r_{N-1};t}} \fintker{r_{N-1}}{t}{z^{N-1}} \overline{\fjack{[-s_{N-1}^{r_{N-1}}]}{t}{z^{N-1}}} \fdjack{\mu_{N-1}}{t}{z^{N-1}} \prod_{i=1}^{r_{N-1}} \frac{\dd z^{N-1}_i}{z^{N-1}_i} \cdot \ket{\theta^+_{\vec{r},\vec{s}}} \notag \\
&=\sum_{\substack{\mu_1,\mu_2,\dots, \mu_{N-1} \\ \hphantom{\mu_1,}	\nu_2,\dots,\nu_{N-1}}} \prod_{k=1}^{N-2} \jprod{\djack{\mu_k}{t}, \jack{\nu_{k+1}+[-s_k^{r_k}]}{t}}{r_k}{t} \cdot \jprod{\djack{\mu_{N-1}}{t}, \jack{[-s_{N-1}^{r_{N-1}}]}{t}}{r_{N-1}}{t} \rho_+ \brac*{\fjack{\mu_1}{t}{y^1}} \prod^{N-1}_{k=2} \rho_+ \brac*{\fjack{\mu_k / \nu_k}{t}{y^k}} \cdot \ket{\theta^+_{\vec{r},\vec{s}}} \notag \\
&= \sum_{\nu_2,\dots,\nu_{N-1}} \brac*{\prod_{k=1}^{N-2} \nc{\nu_{k+1} + [-s_k^{r_k}]}{t}(r_k)\cdot \nc{[-s_{N-1}^{r_{N-1}}]}{t}(r_{N-1})} \notag \\
&\quad \cdot \rho_+ \brac*{\fjack{\nu_2+[-s_1^{r_1}]}{t}{y^1}} \prod^{N-2}_{k=2} \rho_+ \brac*{\fjack{\brac*{\nu_{k+1}+[-s_k^{r_k}]} / \nu_k}{t}{y^k}} \cdot \rho_+ \brac*{\fjack{[-s_{N-1}^{r_{N-1}}] / \nu_{N-1}}{t}{y^{N-1}}} \ket{\theta^+_{\vec{r},\vec{s}}}.
\end{align}
As before, the factor $\nc{[-s_{N-1}^{r_{N-1}}]}{t}(r_{N-1})$ does not depend on the summation indices $\nu_2,\dots,\nu_{N-1}$,
appears in every summand, and is non-zero, so it can be suppressed. Moreover, the skew-Jack polynomials vanish unless the summation indices $\nu_2,\dots,\nu_{N-1}$ satisfy the relations
\begin{align}
\nu_{k} \subseteq \nu_{k+1} + [-s_{k}^{r_{k}}],\quad k=2,\dots,N-2, \quad \nu_{N-1} \subseteq [-s_{N-1}^{r_{N-1}}].
\end{align}
Thus, the singular vector \(\scrs{+}{\vec{r}} \ket{\eta^+_{\vec{r},\vec{s}}} \in \Fock{\theta^+_{\vec{r},\vec{s}}}\) is proportional to
\begin{multline}\label{eq:WNSVFormula}
  \sum_{\nu_2,\dots,\nu_{N-1}} \brac*{\prod_{k=1}^{N-2} \nc{\nu_{k+1} + [-s_k^{r_k}]}{t}(r_k)}
	\rho_+ \brac*{\fjack{\nu_2+[-s_1^{r_1}]}{t}{y^1}} \\
	\cdot \prod^{N-2}_{k=2} \rho_+ \brac*{\fjack{\brac*{\nu_{k+1}+[-s_k^{r_k}]} / \nu_k}{t}{y^k}}
	\cdot \rho_+ \brac*{\fjack{[-s_{N-1}^{r_{N-1}}] / \nu_{N-1}}{t}{y^{N-1}}} \ket{\theta^+_{\vec{r},\vec{s}}}.
\end{multline}
This is our final formula for $\Wvoa{n}$ singular vectors generalising the
$n=3$ case in \eqref{eq:W3SVFormula}.  As before, considering the summand
with $\nu_2 = \cdots = \nu_{N-1} = \sqbrac*{\ }$ shows that the \rhs{} is
non-zero for every \(\vec{r}\in\ZZ_{\ge0}^{N-1}\), \(\vec{s}\in\ZZ_{\le0}^{N-1}\) and \(t\in\CC\setminus \QQ_{\le0}\).  This singular vector formula also has the nice property of being comparatively easy to evaluate using computer algebra packages such as \textsc{SageMath}.

\section*{Acknowledgements}

We thank Peter Forrester and Ole Warnaar for discussions relating to the results presented here.
DR's research is supported by the Australian Research Council Discovery Project DP160101520 and the Australian Research Council Centre of Excellence for Mathematical and Statistical Frontiers CE140100049.
SW's research is supported by the Australian Research Council Discovery Early
Career Researcher Award DE140101825 and the Australian Research Council Discovery Project DP160101520.

\flushleft
\singlespacing


\end{document}